\documentclass[12pt]{article}

\usepackage[utf8]{inputenc} % set input encoding (not needed with XeLaTeX)

\usepackage{geometry} % to change the page dimensions
\usepackage{longtable}
\usepackage{booktabs} % for much better looking tables
\usepackage{array} % for better arrays (eg matrices) in maths
\usepackage{paralist} % very flexible & customisable lists (eg. enumerate/itemize, etc.)
\usepackage{verbatim} % adds environment for commenting out blocks of text & for better verbatim
\usepackage{subfig} % make it possible to include more than one captioned figure/table in a single float

\usepackage{authblk}
\usepackage{amssymb}
\usepackage{amsthm}
\usepackage{amsmath}
\usepackage[pdfstartview=FitH]{hyperref}
\usepackage{graphicx} % support the \includegraphics command and options
\usepackage{parskip}

\usepackage{fancyhdr} % This should be set AFTER setting up the page geometry
\def\tightlist{}

\title{ The Quantitative Finance Aspects of Automated Market Markers in DeFi \\ Version v0.1}
\author{Stefan Loesch \\ \href{mailto:stefan@topaze.blue}{stefan@topaze.blue} }

\affil{  \href{https://topaze.blue}{topaze.blue}  }
\date{January 24, 2022}
\begin{document}

\maketitle

\begin{abstract}
   Automated Market Makers (AMMs) are a class of smart contracts on Ethereum and other blockchains that "make markets" autonomously. In other words, AMMs stand ready to trade with other market participants that interact with them, at the conditions determined by the AMM. In this this paper, which relies on the existing and growing corpus of literature available, we review and present the key mathematical and quantitative finance aspects that underpin their operations, including the interesting relationship between AMMs and derivatives pricing and hedging. 

This paper is a chapter of The AMM Book (theammbook.org) which embeds it into a wider and less technical context, including economics, regulations and a description of the related eco system.
\end{abstract}

\pagebreak

\tableofcontents

\pagebreak

\hypertarget{preface}{%
\section{Preface}\label{preface}}

Automated Market Makers, short AMMs, are smart contracts that
autonomously make markets in tokens on a blockchain, in particular the
Ethereum blockchain. We will in the years to come see a convergence
between traditional finance (``TradFi'') and the emerging decentralized
finance (``DeFi''), and whilst it is too early to understand where it
will end up, it is highly likely that AMMs will play a central role in
this convergence.

AMMs, like trading venues in traditional finance, are places where
assets change owners. That means they are in the center of the financial
system -- without them the financial system could not exist. They are
also highly complex entities, both in their own right, and in their
interaction with other parts of the system. This is the reason why we
are currently working on The AMM Book
(\href{https://theammbook.org}{theammbook.org}) - it is important for
everyone, especially in the TradFi and regulatory community who have not
yet been exposed to the topic, to understand what those AMMs are and how
they work.

The book covers AMM from various different angles -- their technical
implementation, their economics, their regulation, and, last but
certainly not least, their internal mechanics. AMMs are following a
passive trading strategy: they offer to trade with everyone who
approaches them, on the terms determined by their internal algorithms.
It is well known since Black, Scholes and Merton wrote their seminal
papers {[}Black Scholes 73, Merton 73{]} that trading strategies and
financial derivatives are closely related. This suggests -- and it turns
out this is true -- that the quantitative finance apparatus that
underpins modern option pricing theory is very well suited to study
AMMs.

We reserved a chapter in our book describing and reviewing the
quantitative finance aspects of Automated Market Makers. This is a
highly specialst topic that is covered both by industry practitioners
and by academics in the world's leading universities, with those two
groups exhibiting a significant overlap. The primary vehical for
advancing knowledge in that world are peer-reviewed papers. We therefore
decided to publish this chapter independently from the book so that the
community can review it -- and to publish it as early as possible so
that at the time the book is ready for publication the paper has
undergone a thorough review and revision process.

Without further ado, please let us thank you for reading this paper and
please, do contact us with any comments, suggestions and in particular
errors.

\begin{center}\rule{0.5\linewidth}{0.5pt}\end{center}

London, January 2022

Stefan Loesch
(\href{mailto:stefan@topaze.blue}{\nolinkurl{stefan@topaze.blue}})

\hypertarget{introduction}{%
\section{Introduction}\label{introduction}}

An AMM is an automated agent, specifically a smart contract on a
blockchain, that is holding two or more assets, and that is willing to
trade with anyone who matches the AMM's price. In this paper, we only
consider \emph{independent} or \emph{untethered} AMMs, ie AMMs who do
not rely on any external information in their decision making process.

If we place ourselves at a specific point in time, with a specific state
of the AMM (notably, its current asset holdings) then the response of
the AMM is determined \emph{only} by its \emph{response algorithm}.
There are various ways to specify this algorithm, but essentially it
must allow the AMM, and its users, to determine at which effective price
the AMM is willing to trade for every potential transaction that is
presented to it.

To give an example -- and to already introduce some notations we will
use throughout this paper -- we assume the AMM contains two tokens, CSH
and RSK. CSH is the numeraire asset, and RSK is the risk asset. Those
designations are arbitrary and could be swapped, but it is one of the
inconveniences of mathematical finance that for most calculation it is
necessary to choose a numeraire, and this choice is often arbitrary.

There are numerous ways how the response algorithm could be structured,
depending on the practical application. Example include the following

\begin{itemize}
\item
  \textbf{Fixed Price Token Distribution Contract.} A fixed price token
  distribution contract is selling RSK at a fixed price against CSH,
  until it runs out of RSK tokens. At this point it halts, as it is not
  buying RSK at any price.
\item
  \textbf{Increasing Price Token Distribution Contract.} An increasing
  price contract increases the price with the number of tokens sold. If
  this price goes to infinity when the amount of tokens held in the
  contract goes to zero then it is possible to ensure that its token
  supply is never fully depleted. This contract will also only sell RSK
  but never buy it.
\item
  \textbf{Fixed Price Trading Contract.} A fixed price trading contract
  is similar to the corresponding token distribution contract, except
  that it is not only selling RSK, but it is also buying it, at a fixed
  price against CSH. This contract stops trading in one direction if
  runs out of either RSK or CSH. It does not halt however, it will
  always trade in at least one direction.
\item
  \textbf{Increasing Price Trading Contract.} This contract is also
  similar to its corresponding distribution contract, except that it is
  trading both ways. If the trading price of RSK goes to infinity when
  the contract runs out of RSK, and to zero when it runs out of CSH, the
  contract will never run out of tokens, and therefore will always be
  willing to trade in either direction, albeit possibly at prices that
  are unattractive when compared to the market.
\end{itemize}

\hypertarget{general-amm-mathematics}{%
\section{General AMM Mathematics}\label{general-amm-mathematics}}

\hypertarget{key-concepts}{%
\subsection{Key concepts}\label{key-concepts}}

We have seen that an AMM requires a response algorithm to handle the
trades proposed to it. There are a number of different yet ultimately
equivalent ways to formalize this algorithm. The first one is the
\textbf{price response function} (\textbf{PRF}) that we denote
\(\pi(\Delta x)\). The PRF associates a price to every quantity of RSK
that someone wants to sell (\(\Delta x > 0\); AMM buys) or buy
(\(\Delta x < 0\); AMM sells). The price can formally be zero or
infinity respectively if the AMM is not ready to trade.

A market maker is expecting to make money buying and selling, and is
generally doing so by quoting a bid/ask spread or charging a fee.
Economically the impact of both is that the price at which an AMM is
buying is higher than the one at which it is buying. This can be
modelled with the bid/ask spread which corresponds to a discontinuity of
the price function at \(\Delta x = 0\) with

\[
\lim_{\Delta x\to 0+} \pi(\Delta x) - \lim_{\Delta x\to 0-} \pi(\Delta x)= s
\]

where \(s\) is the bid/ask spread. Also generally prices worsen for the
AMM counterparty when the trade gets bigger, therefore

\[
x_1 > x_0 \Rightarrow \pi(x_1) \leq \pi(x_0)\
\]

meaning that \(\pi(x)\) is a decreasing function in the AMM's current
portfolio holdings of RSK, \(x\).

Another way of determining the response is by providing an indifference
curve and a fee curve. The \textbf{indifference curve} \(y(x)\)
determines the states \(x,y\) (quantity of RSK, CSH respectively) which,
ignoring fees, are accessible by trading with the AMM. So assume the AMM
currently holds \(x_0\) of RSK and \(y_0\) of CSH, with
\(y_0 = y(x_0)\). Then, ignoring fees, the AMM is willing to trade
towards any point \(x_1,y_1\) with \(y_1 = y(x_1)\) or better. In other
words, the AMM is willing to engage into an exchange
\(\Delta x = x_1-x_0\) if and only if \(\Delta y = y(x_1)-y(x_0)\) or
better. Note that \(\Delta x\) and \(\Delta y\) will always have
different signs.

We now need to include fees. One way of doing so is with the \textbf{fee
curve} \(\varphi(x, x')\) -- which may simplify to \(\varphi(\Delta x)\)
-- and which corresponds to the amount of CSH that anyone trading with
the AMM must pay over and beyond what is needed to stay on the
indifference curve. In other words

\[
\Delta y_{\mathrm{actual}}(\Delta x) = \Delta y(\Delta x) \pm \varphi(\Delta x)
\]

where the sign \(\pm\) is chosen in a manner that it is of benefit to
the AMM. A common choice is
\(\varphi(\Delta x) \propto \Delta y(\Delta x)\), ie charging a
percentage fee. The fee can either be held within the AMM asset pool, in
which case the pool moves to a different indifference curve.
Alternatively, the fee can be set aside or distributed, in which case
the indifference curve remains the same.

The price function \(\pi(\Delta x)\) can be recovered from the
indifference and fee curves as

\[
\pi_{\mathrm{actual}}(\Delta x) = 
\frac{\Delta y_{\mathrm{actual}}(\Delta x)}{\Delta x} = 
\frac{\Delta y(\Delta x) \pm \varphi(\Delta x)}{\Delta x}
\]

We conjecture that the other way works equally well and is well defined
and unique, assuming we define the indifference curve as not extracting
any fees.

Another, very popular way to determine the indifference curve is by
using a \textbf{characteristic function} \(f(x,y)\). In this case the
indifference curves \(y_k(x)\) are implicitly determined by the
condition

\[
f(x, y) = f(x, y_k(x)) = k
\]

where \(k\) is a constant that identifies a specific indifference curve
within the set of indifference curves described by \(f\).

Finally, another term that is commonly used within the AMM framework is
that of a \textbf{bonding curve}. Unfortunately the meaning of this term
is not always consistent -- it can refer to a PRF (with or without fee
component), an indifference curve, or a characteristic function. In
order to not add to this confusion we will not use the term
\emph{bonding curve} here, but will always use one of the specific terms
we have defined above.

To summarize, above we have introduced three mostly equivalent concepts,

\begin{itemize}
\item
  the \emph{price response function} (\emph{PRF}) which lends itself
  best to economic analysis,
\item
  the \emph{indifference curve} which lends itself best to actual
  implementation of AMMs, and
\item
  the \emph{characteristic function} which is in many ways the most
  elegant of those objects and lends itself best to a mathematical
  analysis.
\end{itemize}

\hypertarget{the-micro-economics-of-the-price-response-function}{%
\subsection{The micro economics of the price response
function}\label{the-micro-economics-of-the-price-response-function}}

\hypertarget{demand-and-supply-curves-in-microeconomics}{%
\subsubsection{Demand and supply curves in
microeconomics}\label{demand-and-supply-curves-in-microeconomics}}

The PRF is closely related to the concept of demand and supply curves at
the base of microeconomic analysis. Before we move on we give a very
brief review of the topic for those who may not be familiar with it. For
a more thorough discussion, see any microeconomics textbook, eg
{[}Pyndick Rubinfeld{]}.

The concept of the \emph{supply curve} is rooted in the \textbf{cost
curves} which originated in commodity markets. The context is that there
are many different producers of a fully interchangeable goods
(``commodity goods''), and they produce those goods at a certain
individual cost. The \emph{cost curve} is the curve that first sorts the
producers by their production cost, and that then plots the produced
quantities on the x axis and the corresponding cost on the y axis. By
construction the cost curve is an upward-sloping step function. In
practice it is often approximated with a continuous function. An example
for a cost curve is shown in the graphics below

\includegraphics[width=12cm,keepaspectratio]{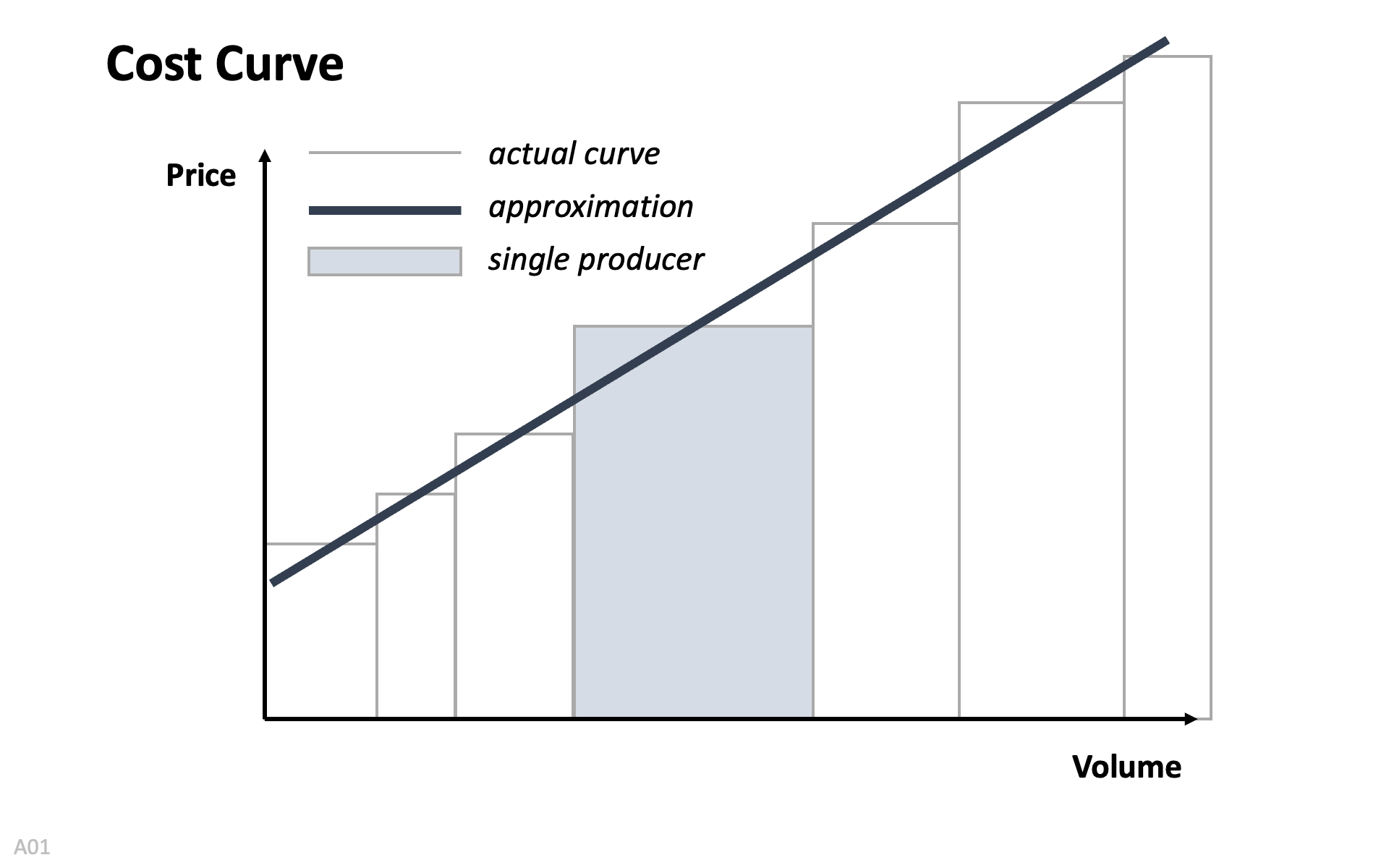}

For simplicity we will ignore fixed costs here and assume all costs are
variable. Also cost can include a \emph{cost of equity}, in other words
a minimum profit margin. With this conditions it is reasonable to assume
that producers are willing to sell whenever the price is above their
individual cost level, and to be content not selling whenever the price
is below. In other words, the \emph{cost curve} turns into the market
\textbf{supply curve}. It is a two-way association, but generally it
serves to associate a supply level with a given price, as shown in the
chart below

\includegraphics[width=12cm,keepaspectratio]{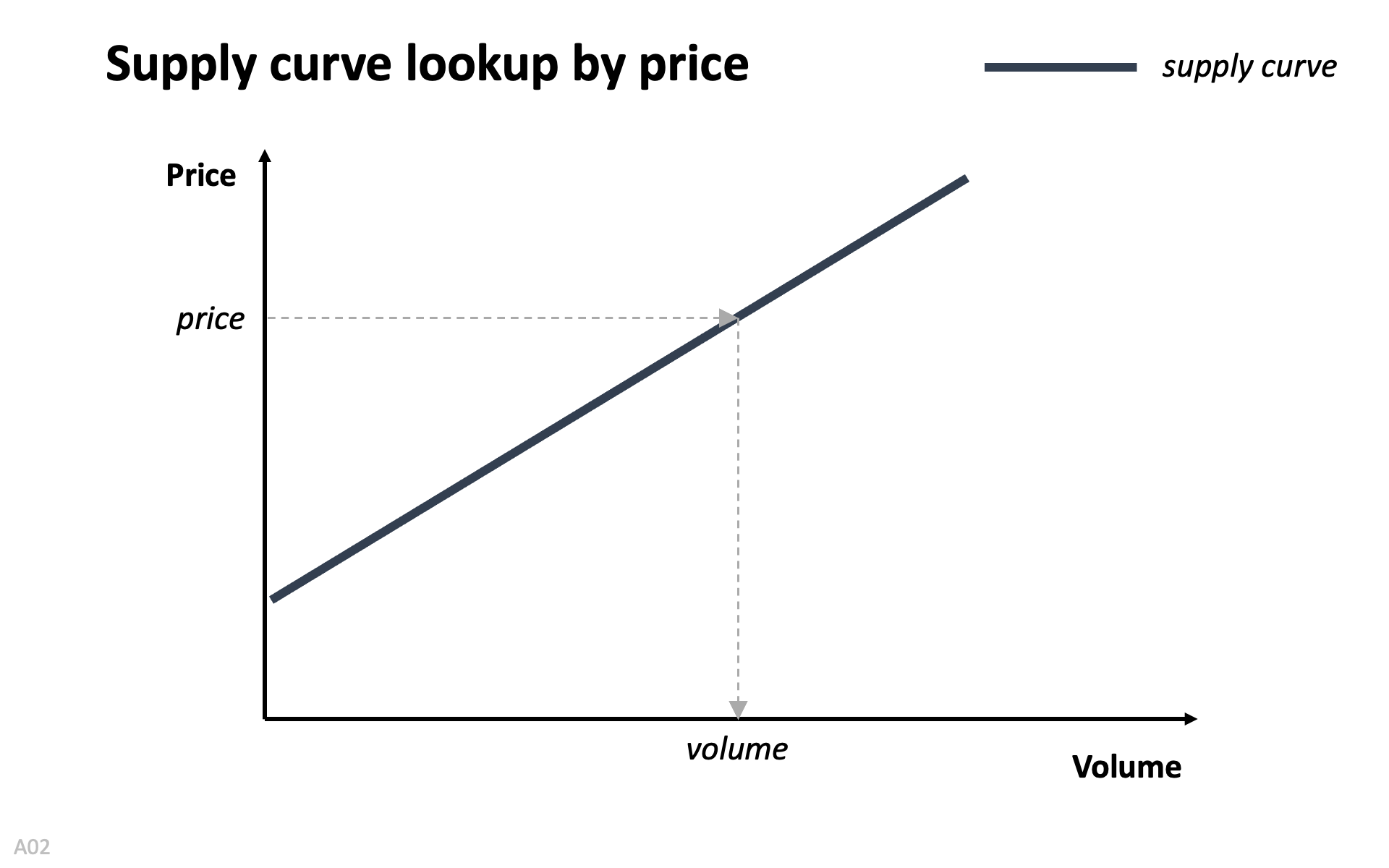}

Complementing the supply curve is the \textbf{demand curve}. It is based
on the same mental model and it constructed similarly. The assumption is
that there are numerous buyers in the market, that they all are willing
to buy below a certain price, and that they are content not buying above
that price. Again, those buyers are sorted by price, this time in
descending price order. The resulting demand curve is a downward-sloping
step function that again is often approximated with a continuous
function.

Combining demand and supply curve in the same diagram allows
determininig the \textbf{equilibrium price}, which is at the price level
where the supply and demand curves meet. All buyers to the left of this
point are, by construction, willing to buy at the equilibrium price or
below, and all sellers to the left are willing to sell at the
equilibrium price or above. By construction, the quantity sold matches
the quantity bought, so everyone to the left of the intersection point
will transact, and everyone to the right of it will not, and all
transactions will happen at the equilibrium price. This price is also
sometimes referred to as \textbf{market clearing price} as this is the
price at which the market clears, ie where no possible transactions are
open. A combined supply and demand curve and determination of the
associated clearing price is shown below.

\includegraphics[width=12cm,keepaspectratio]{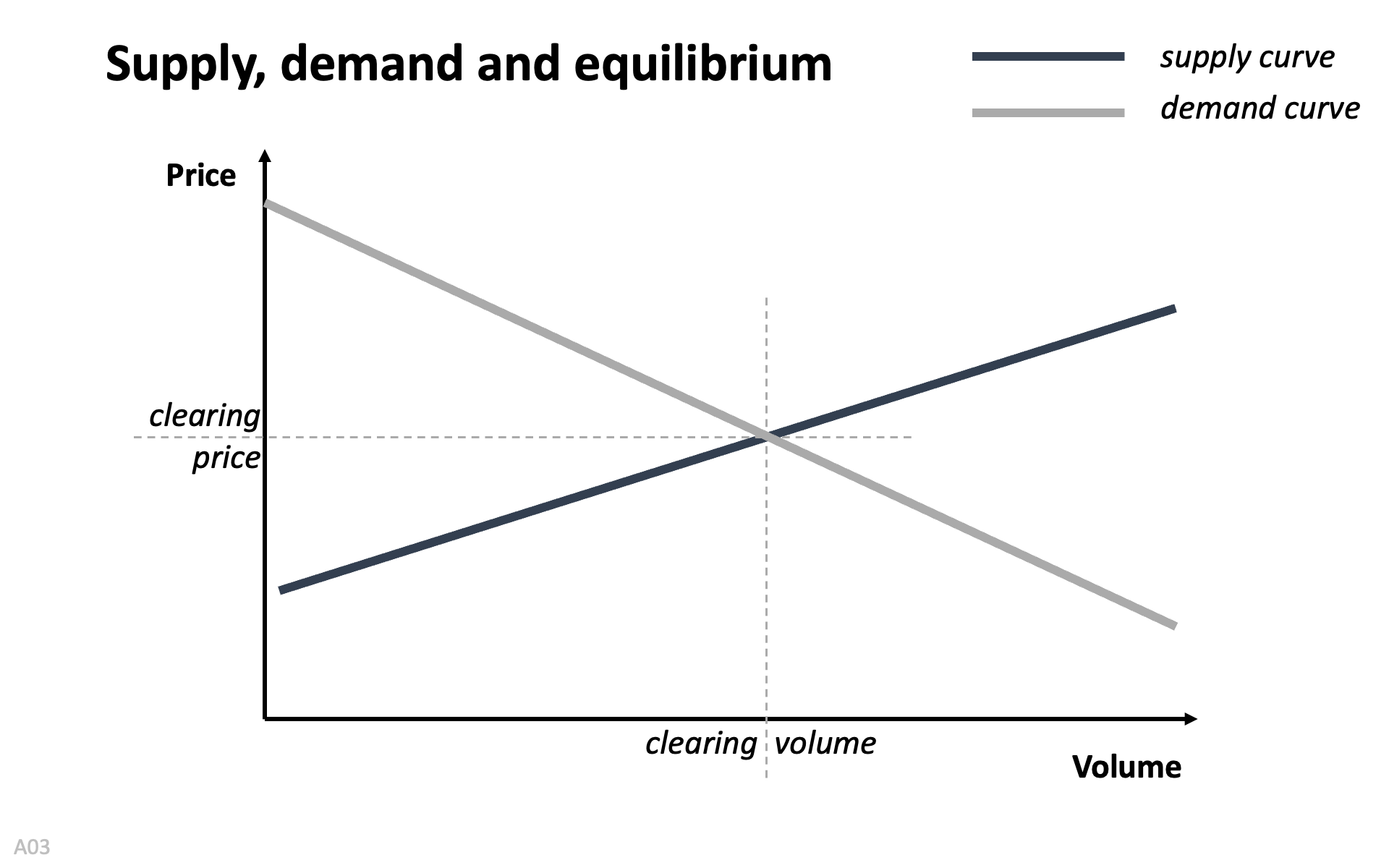}

\hypertarget{supply-and-demand-curves-in-financial-markets}{%
\subsubsection{Supply and demand curves in financial
markets}\label{supply-and-demand-curves-in-financial-markets}}

The concept of supply and demand curves is also useful in financial
markets, provided one understands the mental model the underpins it.

Let's start with an excessively simplistic but nevertheless enlightening
view of the world: we assume that \emph{every} market participant has a
view on \emph{every} asset in the market, in the sense that they have a
view on its fair price. Moreover we assume that market participants will
act on their views, meaning they will be buyers if the asset is
available below what they consider its fair value, and they will be
sellers if someone is willing to pay more.

In a multi-asset world this can get exceedingly complex, so we take
refuge in our two asset world of RSK can CSH. We can then again assemble
the market participants and sort them by their price assessment of RSK
vs CSH. Increasing or decreasing does not matter in this case, and we
choose decreasing. In this case the curve ressembles a demand curve. We
know what the supply curve is: the supply of RSK is fixed, so the supply
curve is a vertical line. The \emph{market clearing price} is where the
vertical line intersects with the demand-like curve we have constructed.
This relationship is shown in the chart below

\includegraphics[width=12cm,keepaspectratio]{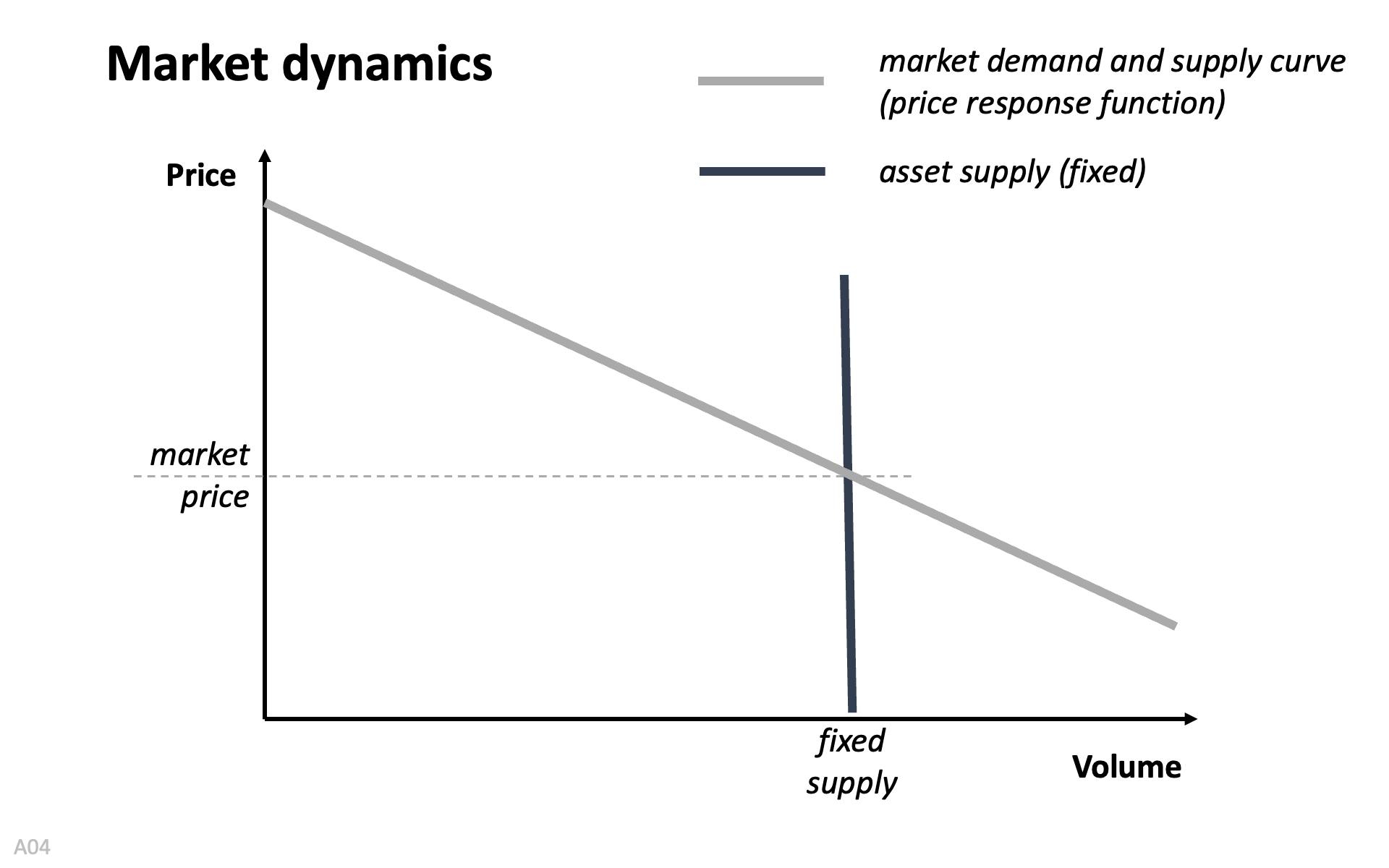}

As stated before, this model is overly simplistic. Most market
participants do not trade that way, and would not even trade that way if
there were zero transaction costs. Most people will hold on to their
investments for a while. At best we can assume that they sell above
certain sell threshold to take profits, and buy below a certain buy
threshold if they think that there is sufficient potential. This
situation can be described nicely with a supply and demand curve: the
demand curve is people adding RSK to their portfolio on the downside,
and the supply curve is people selling RSK on the upside.

However, reality is even more complex. For example, if markets fall,
market participants may want to cut their losses, so they sell on the
downside. Similarly, people may want to buy on the upside for fear of
missing out. Those dynamics do \emph{not} fit into a simple static
supply and demand curve model. At the very least one needs to assume
that the market price dynamics itself feeds back into market supply and
demand curves. This destroys a lot of the simplicity and elegance of
this approach.

\hypertarget{supply-and-demand-curves-in-market-making}{%
\subsubsection{Supply and demand curves in market
making}\label{supply-and-demand-curves-in-market-making}}

We have seen under the previous heading that supply and demand curves
can be applied in markets in general, but that they suffer from some
shortcomings. One area where they work well however is in the analysis
of market makers, ie market participants that stand ready to engage in
trades in case other market participants are willing to meet their
price.

The best manifestation of supply and demand curves is in a market's
order book, more specifically its limit order book (``buy at this price
or below; sell at this price or above''). Those ordder are supply and
demand that will hit the market when prices move. However, the supply
and demand from the order book does not necessarily correspond to the
real short term supply and demand. There are many participants who
monitor the market continously which allows them to add or cancel orders
in reaction to price movements. Therefore the effective market demand
and supply is usually different curve shown in the public order book.

With those caveats out of the way, we remind ourselves of the
\emph{price response functions} (\emph{PRFs}) that we discussed above:
it turns out that is simply describes a static order book. When prices
increase (decrease) the PRF will trigger a known amount of sell (buy)
orders. In other words:

\begin{quote}
an AMM effectively is a static order book.
\end{quote}

\hypertarget{aggregating-prfs}{%
\subsubsection{Aggregating PRFs}\label{aggregating-prfs}}

A market is the superpositon of its participants. In fact, it is a
linear superposition, which allows us to easily aggregate the PRFs of
multiple AMMs. Moreover it allows us to treat every liquidity position
as its own individual micro AMM, greatly simplifying the AMM
mathematics.

Supply and demand curves and PRFs are aggregated along the x-axis, not
the y-axis. Practically speaking this corresponds to putting all
individual positions into a single big pool and sorting them anew by
price. This will interlace the positions coming from the underlying
curves, placing those with similar prices close to each other. An
example for this is shown in the chart below.

\includegraphics[width=12cm,keepaspectratio]{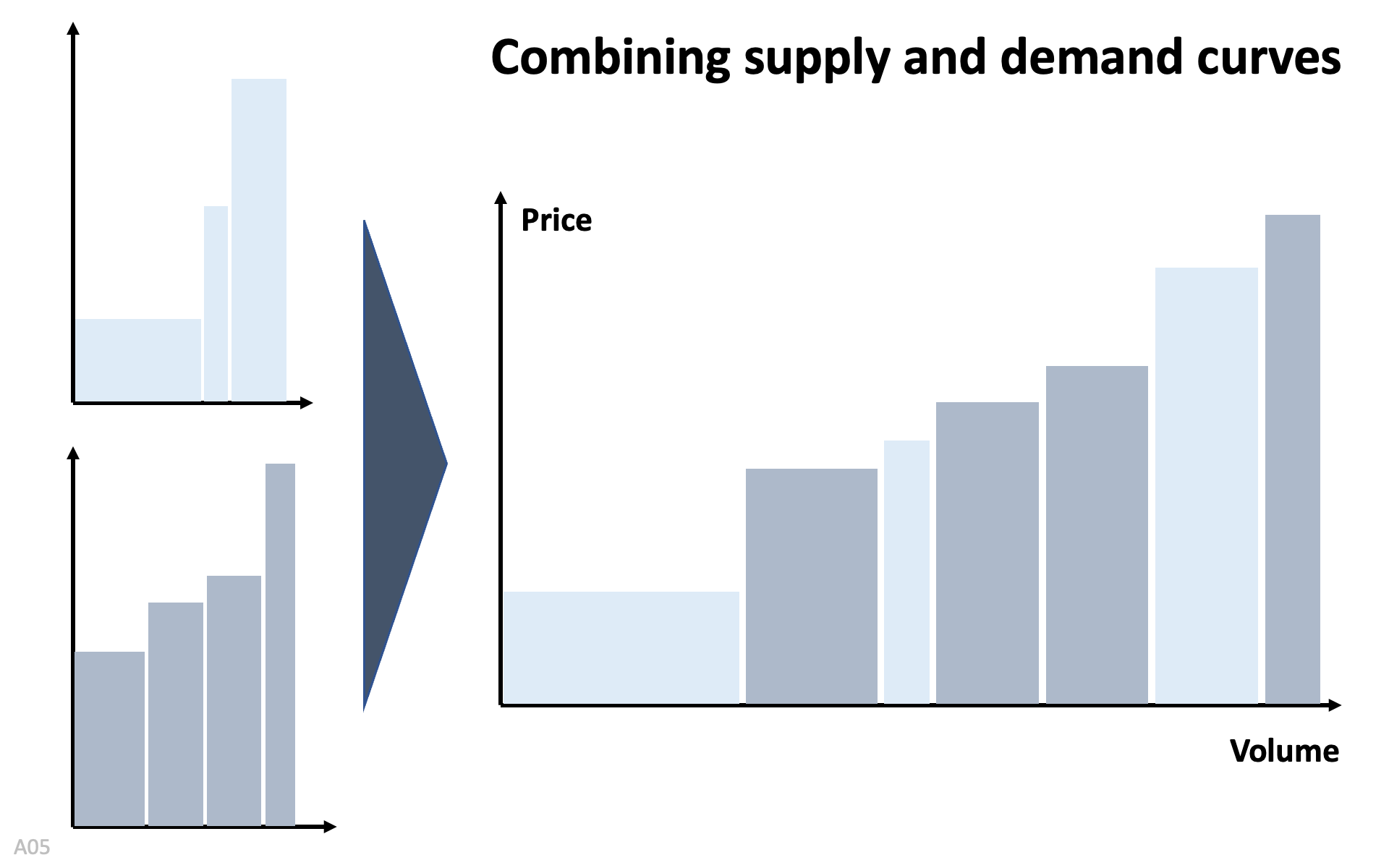}

The continous case is exactly the same: if the have two PRFs
\(\pi_1(x), \pi_2(x)\) then those will get aggregated along the x-axis.
If we denote \(\pi^{-1}\) the inverse function \(\pi\) then the
aggregated PRF is the inverse of the sum of the inverse of the
constituent functions. Written as formula this becomes

\[
\pi(x) = \left(\pi^{-1}_1(x)+ \pi^{-1}_2(x)\right)^{-1}
\]

where contrary to elsewhere in this paper, \(\pi^{-1}\) denotes not
\(1/\pi\) but the inverse function of \(\pi\), associating volume with
price and not vice versa.

\hypertarget{optimal-routing}{%
\subsubsection{Optimal routing}\label{optimal-routing}}

When there are multiple trading venues, traders have to decide where to
route their trades -- this is not different in defi than in traditional
finance. The difference is that trade routing in defi is comparatively
easy as all DEXes expose well-defined APIs, and in many cases they will
even expose the same API.

Naively one may think that trade routing is as easy as sending the trade
to the DEX that currently offers the best price, but this is mistaken.
The reason for this is \emph{slippage}, ie the fact that the price gets
worse (and possibly substantially worse) with an increase in trade size.
Therefore it is often beneficial to split a trade into parts and route
those parts to different DEXes thereby generating less overall slippage
as the aggregate liquidity of the system is used.

We now place ourselves in a world where we have a number of otherwise
identical DEXes serving the same trading pair. They are currently in
equilibrium at the same price level, they charge no fees (or the same
percentage fees), and gas costs are negligible when compared to the
trade volume. It is easy to verify mathematically that in this case the
optimal (slippage minimizing) routing is to split the trade in
proportion to the liquidity depth in the respective pools.

It is equally easy to see this from the previous discussion on PRFs: the
market PRF is the aggregate PRF of the individual DEXes, and best
pricing will be achieved when the liquidity is used in the order in
which it appears in the PRF, best price to worst. As we've seen in the
previous chart this will usually involve tapping into liquidity provided
by more than one exchange, and therefore splitting up the trade and
routing it accordingly.

This last algorithm holds more generally. To route trades optimally we
can first create the market PRF, remembering for every slice of
liquidity in that PRF where it comes from, eg by coloring it. From the
market PRF we can then see what price we get at which trade volume, and
by looking at the colors of the segments covered we know how to route
the trade. This algorithm does work in presence of percentage fees as
those can simply be converted into a price adjustment. It does however
not work for per-trade fees or gas costs. In case those are relevant --
usually they are not -- it is possible to include them in a numerical
optimization algorithm.

\hypertarget{amm-characteristic-functions}{%
\subsection{AMM characteristic
functions}\label{amm-characteristic-functions}}

The second topic we need to discuss in more detail is that of
\emph{characteristic functions}. As a reminder, a characteristic
function \(f(x,y)\) determines a series of indifference curves
\(y_k(x)\) with the help of the condition \(f(x, y_k(x))=k\).

\hypertarget{scaling-symmetry}{%
\subsubsection{Scaling symmetry}\label{scaling-symmetry}}

The first point to make is that the characteristic functions preserve
the symmetry of the underlying situation, ie there is no
numeraire-related symmetry break which in finance often muddies the
waters. For example, if we look at the indifference function \(y(x)\),
then we have implicitly designated \(y\) as the numeraire asset and
\(x\) as the risk asset, and what happens at the upside when the price
of the risk asset goes to infinity looks different to the downside when
it goes to zero, even though because of the underlying symmetry both
situations are exactly the same.

To make things more concrete, let's peek ahead and introduce the most
well known characteristic function we'll discuss below -- the
\textbf{constant product} function \(f(x,y) = x*y\) and its indifference
function \(y_k(x) = k/x\). As before, \(x\) represents the number of RSK
tokens, and \(y\) the number of CSH tokens, both in their own native
units.

As both \(x\) and \(y\) use their own native tokens we need to analyze
what happens if they re-denominate. Generally in finance, redenomination
applied to everything should not change anything in the real world. This
is not a very deep result -- it simply means that it should not matter
if we record prices in dollars, cents, or millions of dollars.

In particular, if we transform all our denominations the same way -- say
instead of using USD and EUR we use USD cents and EUR cents, and we
simularly add to decimals to all other denominations -- then nothing
should change at all. Again, this is not deep -- it simply means that
the price of a EUR in USD terms is the same as the price of a EUR cent
in terms of USD cents.

Mathematically this means we have a representation of the multiplative
group of positive numbers \(\lambda \in R^+\) on our state space that
acts according to \(\lambda: (x,y) \rightarrow (\lambda x, \lambda y)\),
and we want to understand the implied action on on our characteristic
function. Generally, interesting objects in finance transform under this
symmetry in one of two ways

\begin{enumerate}
\def\labelenumi{\arabic{enumi}.}
\item
  They are \emph{invariant} which means that
  \(f(\lambda x, \lambda y) = f(x, y)\)
\item
  They transform \emph{linearly} (or \emph{homogenously of order 1} but
  this is a mouthful) which means that
  \(f(\lambda x, \lambda y) = \lambda f(x, y)\)
\end{enumerate}

Exchange ratios -- ie the price of one asset expressed in terms of
another asset -- are an example of \emph{invariant objects}. Quantities
on the other hand are \emph{linear objects}. Again, this is not deep:
the first statement says that I can look at EUR and USD or EUR cent and
USD cent. The second one says that when I look at cent instead of EUR
and USD all EUR and USD related quantities get multiplied by 100.

When we look at \(f(x,y) = x*y\) we see that it is neither linear nor
invariant, so maybe it is not ideal. We will park this point for now and
come back to it in a moment.

\hypertarget{transformations}{%
\subsubsection{Transformations}\label{transformations}}

The second point to make is that, if we are only interested in the
indifference curves \(y_k(x)\) determined by \(f(x,y)=k\), then our
problem is overdetermined, in the sense that there will be other
characteristic functions \(\bar{f}\) that yield the same set of
indifference curves.

Let's consider a bijective and suitably regular function
\(h: R \rightarrow R\). It is easy to see that the \(y_k(x)\) implied by
\(f(x,y)=k\) are exactly the same as the \(y_{h(k)}(x)\) implied by
\(h(f(x,y))=h(k)\). In other words: if \(f\) is a characteristic
function, then the composite function \(f_h = h \circ f\) is also a
characteristic function. Provided the \(k\) are transformed accordingly
it is fully equivalent to \(f\).

Coming back to \(f(x,y) = x*y\) we see that if we use
\(h(\kappa)=\sqrt{\kappa}\) then
\(\bar{f}(x,y) = f_h(x,y) = \sqrt{x*y}\). It is easy to verify that in
this case our characteristic function transforms linearly, ie

\[
\bar{f}(\lambda x,\lambda y) = \lambda \bar{f}(x,y)
\]

This in turn suggests that the quantity \(\bar{k}\) with
\(\bar{k} = \bar{f}(x,y)\) may be financially meaningful. It turns out
that it is: \(\bar{k}\) is a measure of the pool size that, contrary to
its total monetary value, is invariant under changes in relative prices.
In other words: because \(\bar{k}\) is the pool invariant, if we ignore
fees it does not change when someone is trading against the pool. An
increase in \(\bar{k}\) therefore indicates either an addition of
liquidity or a pool profit (eg because of fees earned), and an decrease
either a removal of liquidity of losses due to leakage or exploits.

\hypertarget{conditions-characteristic-functions-must-satisfy}{%
\subsubsection{Conditions characteristic functions must
satisfy}\label{conditions-characteristic-functions-must-satisfy}}

We are now analysing what conditions a function \(f\) must satisfy to be
a valid characteristic function. We recall from above that the price
response function is \(\pi(x) = -dy/dx\). Prices must be positive,
therefore we find \(dy/dx < 0\). Moreover, \(\pi(x)\) must be
downward-sloping -- the AMM will buy the risk asset when its price
falls, not sell it and vice versa. Therefore we need to have
\(d\pi/dx\geq 0\) and hence \(d^2y/dx^2 < 0\).

We know that \(df=\partial_x f dx + \partial_y f dy\) where \(\partial\)
denotes the partial derivative, and that along the indifference curves
we must have \(df=0\) by construction. Rearranging those terms yields

\[
-\frac{dy}{dx} = \frac{\partial_x f}{\partial_y f} > 0
\]

(note the reversal of x and y). This condition essentially states that
the two partial derivatives of \(f\) must have the same sign. In other
words, the gradient vector must either point top-right (between North
and East) or bottom-left (between South and West).

\hypertarget{the-mathematics-of-multi-asset-pools}{%
\subsection{The mathematics of multi-asset
pools}\label{the-mathematics-of-multi-asset-pools}}

As the name implies, multi-asset pools contain more than two assets, and
are prepared to engage into any cross-trades natively. The alternative
to this is a network of connected two-asset pools. Those can either
follow a \emph{hub-and-spoke design} with one common central asset, or a
\emph{point-to-point design} where crosses have their own pools.

\hypertarget{definitions}{%
\subsubsection{Definitions}\label{definitions}}

A multi-asset pool of \(N+1\) assets is most easily defined by its
characteristic function \(f(x_0, x_1, \ldots, x_N)\) where the \(x_i\)
are the token amounts in their native denomination. As before we will,
for demonstration purposes, look at a specific characteristic function

\[
f(x_0, x_1, \ldots, x_N) = x_0 \cdot x_1 \cdot \cdots \cdot x_N
\]

We should note that the better function is the geometric average

\[
\bar f(x_0, x_1, \ldots, x_N) = \sqrt[N+1]{x_0 \cdot x_1 \cdot \cdots \cdot x_N}
\]

but we know from the symmetry discussion above that the two functions
above are equivalent, and \(f\) is easier to deal with than \(\bar f\).
We, in this section only, also adopt the convention that \(x\) without
an index represents the entire vector

\[
x \equiv (x_0, x_1, \ldots, x_N)
\]

which allows us to abbreviate our characteristic function as \(f(x)\).

We can choose a numeraire asset if we want -- in which case we choose
\(x_0\) -- or we can treat the whole problem as a symmetrical problem
where all assets are considered \emph{risky}.

We need exhange ratios \(\pi_{ij}(x)\) for each of the pairs
\(x_i,x_j\), and we adopt the convention that the second index (\(j\) in
this case) is the numeraire. The \(\pi_{ij}(x)\) are determined by
\emph{partial} derivatives.

\[
\pi_{ij}(x) = \frac{\partial_j f(x)}{\partial_i f(x)}
\]

where we use the shortcut notation
\(\partial_i \equiv \partial / \partial x_i\). Note that the numeraire
index is in the numerator.

For convenience we also define the single-index functions
\(\pi_i\equiv\pi_{i0}\), so if the second index is missing the numeraire
is implied, and we have

\[
\pi_{i}(x) = \frac{\partial_0 f(x)}{\partial_i f(x)}
\]

\hypertarget{consistency-and-geometry}{%
\subsubsection{Consistency and
geometry}\label{consistency-and-geometry}}

When we have a system of prices, those price systems must be arbitrage
free. Firstly, the price for the reverse exchange must be the inverse of
the price, ie

\[
\pi_{ij}=\frac{1}{\pi_{ji}}
\]

Secondly, the exchange ratio of a direct exchange between \(x_i, x_k\)
must be the same as the exchange going via \(x_j\), therefore

\[
\pi_{ik} = \pi_{ij}\cdot\pi_{jk}
\]

Both of those conditions can be easily verified going back to the
definitions of the \(\pi_{ij}\). We note that this means that the
\(\pi_i\) are sufficient to span the entire price space via the
relationship

\[
\pi_{ij} = \pi_{i0}\cdot\pi_{0j} = \frac{\pi_i}{\pi_j}
\]

This of course is nothing but the well known fact that, in an arbitrage
free system, it is possibly to choose a numeraire, and once all prices
in the numeraire are fixed all cross exchange ratios are fixed as well.

Geometrically we can think of what we called the invariance curve in two
dimensions as an invariance hypersurface (technically, a codimension-1
manifold embedded into \(R^{N+1}\)). This hypersurface (which from now
on we will simply refer to as indifference ``surface'') is defined by

\[
f(x) = k
\]

or, equivalently, by the differential condition

\[
df(x) = \sum \partial_i f(x) dx_i = 0
\]

Like in the two-dimensional case, we need prices to be positive, ie
\(\partial_i f(x)>0\). In other words, like in the two-dimensional case,
the gradient vector
\(\nabla f(x) = (\partial_0, \partial_1, \ldots, \partial_N) f(x)\) must
point into the first (``top-right'') quadrant of \(R^{N+1}\), ie all
vector components must be positive (we ignore the all-negative case
here; we can use \(-f\) instead of \(f\) if need be).

We also have the second-derivative condition \(\partial_i^2 f(x)>0\)
which we need to ensure that the lower the price of an asset the more of
it the AMM holds and vice versa. Our conjecture is that a \emph{concave}
surface (together with the \emph{``gradient in first quadrant''}
condition) is necessary and sufficient for the function \(f\) to be a
valid AMM characteristic function. In a more pedestrian and not
coordinate free approach we can calculate the derivative of
\(\pi_{ij} = \partial_j f / \partial_i f\) and obtain

\[
\partial_i \partial_i f(x) \cdot \partial_j f(x)
-
 \partial_i f(x) \cdot \partial_i \partial_j f(x) 
 > 0
\]

For our example characteristic function \(f(x) = \prod_i x_i\) we find
that \(\partial_i f(x) = \prod_{i\neq j} x_j\) which is positive as all
\(x_i >0\). Similarly we have all second derivatives
\(\partial_i \partial_j f(x) > 0\) as well.

\hypertarget{amms-and-financial-derivatives}{%
\section{AMMs and financial
derivatives}\label{amms-and-financial-derivatives}}

\hypertarget{black-scholes-and-derivatives-pricing}{%
\subsection{Black Scholes and derivatives
pricing}\label{black-scholes-and-derivatives-pricing}}

Before we look at the relationship between AMMs and derivatives, here a
brief reminder of the key elements of derivatives pricing and hedging
that will be important in what follows. For more details see {[}Hull{]}
or any other option pricing text book.

We start with the \textbf{Black Scholes PDE} (Black Scholes partial
differential equation) which reads

\[
\frac{\partial \nu}{\partial t}
=
-\frac{1}{2}\sigma^2\xi^2\frac{\partial^2 \nu}{\partial \xi^2}
- (r-d)\xi\frac{\partial \nu}{\partial \xi}
+ r \nu
\]

Here \(\nu\) is the price of the derivative, and \(\xi\) is the price of
the underlying. We should point out that the fact that we use the same
symbols \(\nu, \xi\) the we will use in the forthcoming sections is not
by chance. Furtermore, \(r\) is the numeraire funding and deposit rate,
and \(d\) is the dividend yield (in case of equity derivatives) or the
foreign asset funding and deposit rate (in case of fx derivatives).
Finally, \(\sigma\) is the lognormal volatility of the underlying
\(\xi\), and \(\sigma^2\) is the variance.

When working with the Black Scholes equation, and in option pricing more
generally, it is customary to define the \textbf{Greeks}, ie variables
that have a standard meaning. We start with \textbf{Theta} which is
defined as

\[
\Theta = \frac{\partial \nu}{\partial t}
\]

Theta is the time decay or, in other words, the fair option premium
accruing over an infinitesimal time period.

Next one up is \textbf{Delta} which is defined as

\[
\Delta = \frac{\partial \nu}{\partial \xi}
\]

Delta is the \emph{hedge ratio}, ie it determines how much of the
underlying needs to be purchased in a \emph{delta hedge}. Delta is
denominated in units of the risk asset. As \(\xi\) is the price,
\(\xi\Delta\) is the hedge portfolio denominated in the numeraire asset.
The quantity \(\xi\Delta\) is referred to as \textbf{Cash Delta}.

Finally we are looking at \textbf{Gamma} which is defined as \[
\Gamma = \frac{\partial^2 \nu}{\partial \xi^2}
\] The Gamma is the change of Delta with respect to changes in the
price, and therefore indicates how the hedge portfolio must be adjusted
when prices change. Gamma has impractical units, and it is often more
intuitive working with the \textbf{Cash Gamma} which is defined as
\(\xi^2\Gamma\) and which, as the name implies, is also denominated in
the numeraire asset.

Using those Greeks we can rewrite the Black Scholes PDE as

\[
\Theta
=
-\frac{1}{2}\sigma^2\Gamma_{\mathrm{cash}}
- (r-d)\Delta_{\mathrm{cash}}
+ r \nu
\]

The operative part of the Black Scholes equation as far as option
valuation is concerned is only the first term

\[
\frac{\partial \nu}{\partial t}
=
-\frac{1}{2}\sigma^2\xi^2\frac{\partial^2 \nu}{\partial \xi^2}
\]

which we can rewrite with the Greeks as follows

\[
\Theta = 
-\frac{1}{2}\sigma^2\Gamma_{\mathrm{cash}}
\]

The second (``Delta'') term (with \(r-d\)) is the cost of carrying the
hedge, and the third one is about carrying the premium received. We will
ignore those terms in what follows because (a) they are not particular
enlightenting and make the formulas more complex, (b) funding and
deposit rates on crypto asset are a problem to start with and in any
case they are not the same, and (c) we can always place ourselves at the
forward horizon which gets rid of those terms in their entirety.

In other words

\begin{quote}
The value of an option is the product of the variance and the option
Gamma, and the variance is the cost of carrying Gamma across time.
\end{quote}

There are two ways how we can understand the option value. The first one
is by looking at what happens to the \emph{value} of a hedged position
when the spot \(\xi\) moves. The delta hedge has, by design, removed the
component that is linear in \(\xi\), so what is left are the quadratic
and higher order terms. Gamma is the second derivative of the value
function, therefore the second order term for a move of size
\(\sigma \xi\) happening during a time period \(dt\) is
\(\sigma^2 \xi^2 \Gamma\). The option value is simply the sum (integral)
over all those infinitesimal moves.

The second way to understand option value is via the actual hedge.
Remember that Delta is the hedge ratio, and Gamma is the change in Delta
when prices move. If Gamma is \emph{positive} we \emph{buy on the
upside} and \emph{sell on the downside}. This means the hedge
\emph{loses} us money, which means we \emph{receive} an option premium
to compensate for that. Vice versa, if Gamma is negative we \emph{buy
low, sell high} which \emph{makes} us money, so we \emph{pay} an option
premium. This latter description of the option value will become very
important for AMMs which, like some option strategies, will purchase
assets on the upside and sell them on the downside, albeit with a twist.

\hypertarget{calls-puts-and-european-profile-matching}{%
\subsection{Calls, puts and European profile
matching}\label{calls-puts-and-european-profile-matching}}

A \textbf{European Call Option} is the right to buy an asset at a fixed
date in the future \(T\) at a pre-determined \emph{strike price} \(K\).
A \textbf{European Put Option} is the right to sell. It is easy to see
that at maturity we have the following payoffs

\[
C = max(\xi-K, 0)
\] \[
P = max(K-\xi, 0)
\]

A \textbf{Forward} is the right and obligation to buy or sell. We can
easily see that call/put parity holds, meaning that long a call and
short a put is equivalent to being long a forward (all three with the
same strike and notional):

\[
 C-P = \xi-K = F
 \]

Black and Scholes were able to solve the above PDE for calls and puts,
the solution for a call being

\[
C = df \times\left( FN(d_+)-KN(d_-)\right)
\]

where \(df=e^{-rT}\) is the discount factor, \(F=\xi_0*e^{(r-d)T}\) is
the forward, \(N\) is the cumulative standard Normal distribution, \[
d_+ =
\frac{1}{\sigma\sqrt{T}}\left[\ln\left(\frac{F}{K}\right) + \frac{1}{2}\sigma^2T\right]
\]

is one of the ``d'', and

\[
d_- = d_+ - \sigma\sqrt{T}
\]

is the other one.

A \textbf{European profile} is a a function \(\nu(\xi)\) that determines
the payoff at time \(T\). We assume that it is differentiable at least
twice for ease of presentation, but when working with Dirac delta
functions and similar constructs this can easily be extended to
functions with a finite number of discontinuities.

We now want to decompose a European profile \(\nu(\xi)\) into a
portfolio of call options. We denote \(\nu''(\xi)\) the second
derivative of \(\nu\) with respect to \(\xi\) (its Gamma) and we find

\[
\nu(\xi) \simeq \int C_K(\xi)\ \nu''(K)\ dK
 \]

We have used a \(\simeq\) symbol here because the above only matches
from the second derivative onwards (proof by deriving twice respect to
\(\xi\) which transforms the call profile into a Dirac delta).
Technically we can define an equivalence relation \(\simeq\) such that
two functions \(\nu_1,\nu_2\) are equivalent whenever the only different
by an affine function, ie

\[
\nu_1 \simeq \nu_2 \Longleftrightarrow \exists a,b\ \forall \xi: \nu_1(\xi) - \nu_2(\xi) = a\xi+b 
\]

This equivalence class could be called the ``(Delta) hedged payoff
profiles'' because two profiles that after hedging look the same are in
the same class (\(a\) is the Delta and \(b\) is the funding). Note that
this class is of extreme practical importance when managing an option
book because it abstracts from calls and puts (which, because of
call/put parity are the same once hedged). It is the basis for the
so-called ``strike report'' which helps a trader to understand the
convexity (and therefore Gamma bleed) of their position.

\hypertarget{the-strike-density-function-and-the-cash-gamma}{%
\subsubsection{The strike density function and the Cash
Gamma}\label{the-strike-density-function-and-the-cash-gamma}}

If we have a European profile \(\nu(\xi)\) then we define its
\textbf{strike density function} \(\mu(K)\)

\[
\bar\mu(K) = \nu''(K)
\]

It is well-defined on the above equivalence classes, ie if
\(\nu_1,\nu_2\) are in the same class then the associated
\(\bar\mu=\bar\mu_1=\bar\mu_2\) is the same. The object \(\bar\mu\) is a
density, so it usually makes sense to look at the associated
differential form \(\bar\mu(K)dK\). The unit of \(\nu\) is CSH. The unit
of \(\bar\mu(K)dK\) is RSK which we can either obtain by formal
calculation, or by noting that ``1 call'' brings exactly one unit of
Delta, and Delta is measured in RSK. Like we have a \emph{Cash Delta} to
complement the Delta we also have a \textbf{cash strike density
function} \(\mu(K) = K \bar\mu(K)\), ie

\[
\mu(K) = K \nu''(K)
\]

Again we look at the differential form \(\mu(K)\ dK\) which is now
denoted in CSH. Both \(\mu(K)\) and \(\bar\mu(K)\) are closely related
to the Cash Gamma

\[
\Gamma_{\mathrm{cash}}(K) = K \mu(K) = K^2 \bar\mu(K)
\]

\hypertarget{power-law-profiles-under-black-scholes}{%
\subsection{Power law profiles under Black
Scholes}\label{power-law-profiles-under-black-scholes}}

We now look at a specific class of payoff profiles, the \textbf{power
law profiles}, ie profiles of the form \(\nu_\alpha(\xi) = \xi^\alpha\).
Those are particularly easy to deal with under the Black Scholes
equation because they are eigenvectors of the Black Scholes operator.
What this means is the following: the two non-trivial spatial operators
in the Black Scholes equation are \(\xi\frac{\partial}{\partial \xi}\)
and \(\xi^2\frac{\partial^2}{\partial \xi^2}\). As a simple calculation
shows, applying those to the \(\nu_\alpha\) yields

\[
\xi\frac{\partial}{\partial \xi} \nu_\alpha
=
\alpha \nu_\alpha 
\]

and

\[
\xi^2\frac{\partial^2}{\partial \xi^2} \nu_\alpha
=
\alpha (\alpha-1) \nu_\alpha 
\]

so the Black Scholes equation becomes

\[
\frac{\partial}{\partial t} \nu_\alpha
=
\left(
-\frac{1}{2}\sigma^2 \alpha (\alpha-1)
- (r-d)\alpha 
+ r 
\right)
\nu_\alpha 
\]

If we denote the term in parantheses on the right
\(\rho(\alpha)=1/\tau(\alpha)\) then the equation becomes the well-known
equation \(\partial_t \nu_\alpha = \nu_\alpha / \tau(\alpha)\), and we
know the solution to this equation is

\[
\nu_\alpha(\xi, t) = 
e^{t/\tau(\alpha)}\nu_\alpha(\xi, t=0) = 
e^{\rho(\alpha)\ t)}\nu_\alpha(\xi, t=0)
\]

ie an exponential growth of the function over time that is preserving
its shape. As shown above, instead of the characteristic period \(\tau\)
we can also use the exponential growth rath \(\rho = 1/\tau\).

If we chose \(\alpha=0.5\), ie \(\nu_\alpha=\sqrt{x}\), and we as
discussed above assume a vanishing \(r,d\) then

\[
\tau_{0.5} = \frac{8}{\sigma^2}, \ \ 
\rho_{0.5} = \frac{\sigma^2} 8
\]

\hypertarget{analysing-the-constant-product-amm-as-a-financial-derivative}{%
\subsection{Analysing the constant product AMM as a financial
derivative}\label{analysing-the-constant-product-amm-as-a-financial-derivative}}

In this section we analyze the constant product AMM within a Black
Scholes and financial derivatives framework. We want this section to be
self-contained, so we may repeat some arguments that we have made
elsewhere in this paper.

The first thing to understand is that an AMM is effectively an
``investment vehicle'' following a particular trading strategy. This
strategy is not self-financing as it ``hands over'' some of its proceeds
to the arbitrageurs. When looking at the consolidated position of an AMM
and its arbitrageurs the trading strategy is self financing however, and
therefore out general quantitative finance frameworks apply. We simply
have to ensure to split the AMM component from the arbitrageur component
at the end.

Assuming efficient markets, the constant product AMM at every point in
time will have 50\% of its value locked in the risk asset, and the other
50\% in the numeraire asset. This can be easily shown using its
indifference function \(y=k/x\) and the price function \(\pi(x)=k/x^2\):
if we multiply those we find that \(\pi(x)\cdot x=y\), the left hand
side being the value of the risk asset, and the right hand side being
that of the numeraire asset.

We now recall from the previous section that the square root profile
retains its form and grows exponantially with a rate
\(\rho = \sigma^2/8\) when going forwards in time. We can also easily
calculate the cash Delta of that profile as

\[
\Delta_{\mathrm{cash}} = \xi \frac{d \nu}{d \xi}=\frac{1}{2}\sqrt{\xi}=\frac{1}{2}\nu(\xi)
\]

This equation shows that the replication strategy of the square root
profile (its ``delta hedge'') keeps 50\% of the portfolio value in the
risk asset. Therefore 50\% are in the numeraire, therefore they both are
of equal value at the same time, and therefore the constant product AMM
strategy corresponds to a square root value function.

To summarize, we have shown that the particular trading strategy an AMM
follows should lead to the following result when moving forward in time

\[
\nu^*(\xi, t) = \exp(\frac{\sigma ^2}{8}t)\ \sqrt \xi
\]

In reality however the time evolution is as follows

\[
\nu(\xi, t) = \sqrt \xi
\]

the difference being the funds that are lost to arbitrageurs: When we
analyse the trading strategy of an AMM then we see that if the price
moves from \(\xi_0\) to \(\xi_1\) then the AMM allows arbitrageurs to
trade at the geometric average of the prices, \(\sqrt{\xi_0 \xi_1}\).
This price is the same on the way up as it is on the way down, which
proves that, ignoring fees, and AMM hands over \emph{all} Gamma gains to
the arbitrageurs. That's what we see above: the factor
\(\exp(\frac{\sigma ^2}{8}t)\), corresponding to a growth rate of
\(\frac{\sigma ^2}{8}\), is entirely lost for the AMM and handed over to
arbitrageurs instead.

We have drawn a few charts that illustrate this evolution. The first one
is the square root profile over time at different vols. The grey line is
the initial profile at \(t=0\), and the blue and orange lines are the
profiles after 1 year, at 75\% and 150\% vol respectively. The
difference between the grey and the colored lines is what is being
handed over to arbitrageurs (ignoring fees), and we see that for vols
beyond 100\% this can become substantial.

\includegraphics[width=12cm,keepaspectratio]{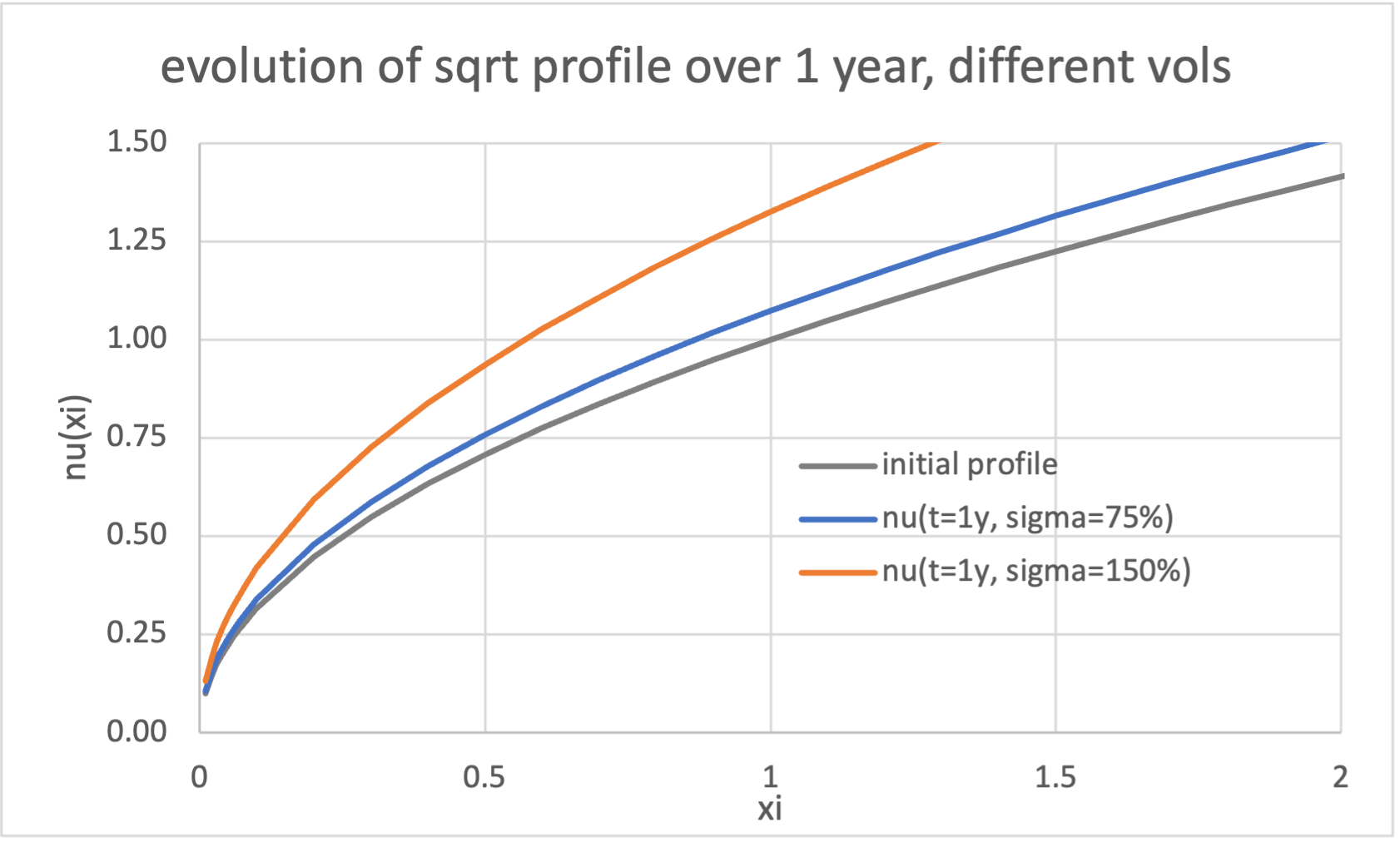}

Now instead of looking at a cross section at different points in time,
we transport a single point through time The x axis is the time in
years, and the y axis is the growth factor. We see nothing much
happening at 50\% vol, but at from 100\% vol onwards the growth becomes
substantial, and very big beyond 150\% vol.

\includegraphics[width=12cm,keepaspectratio]{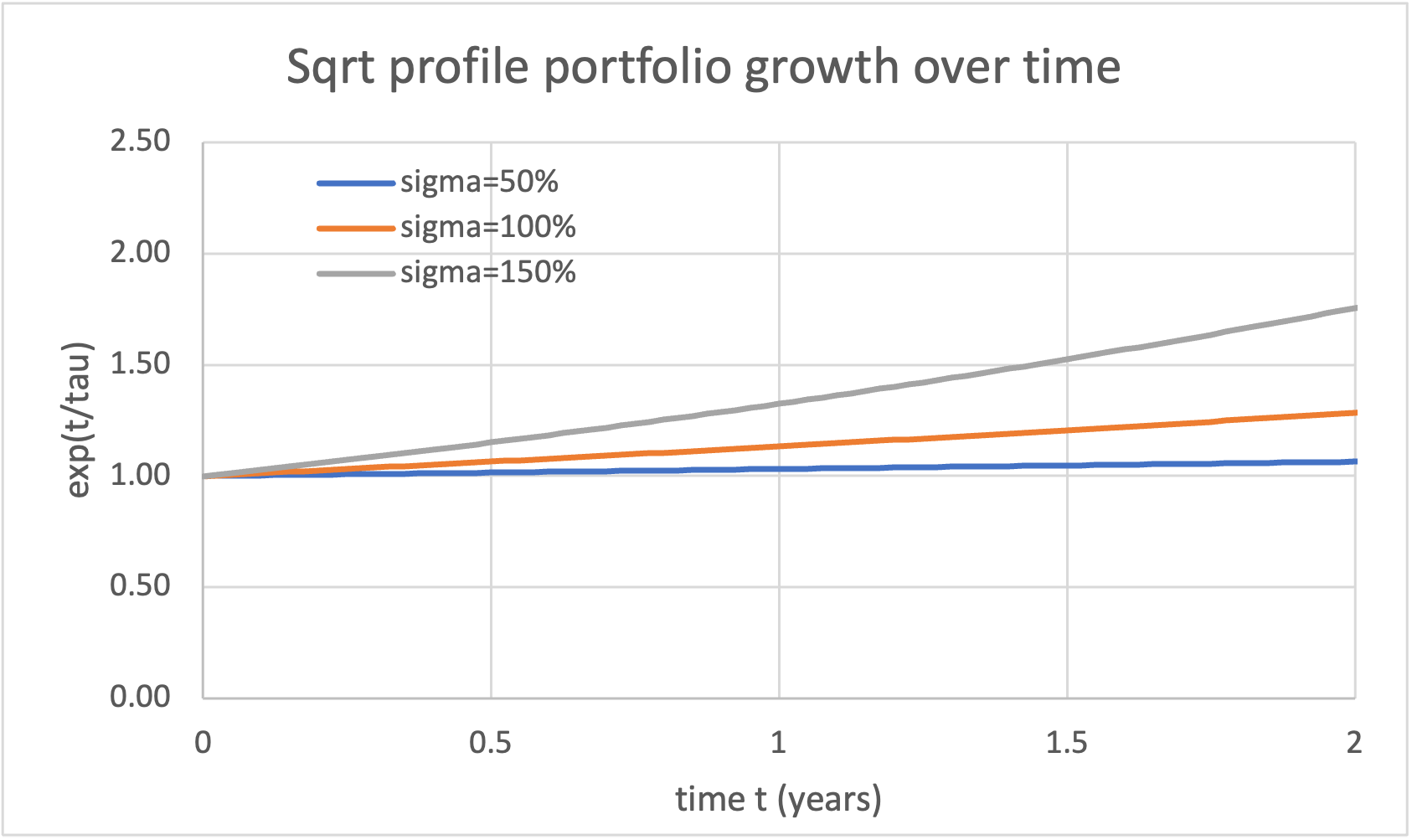}

The final chart here gives as a feel for how the vol impacts growth. On
the x axis we have the volatility \(\sigma\). The blue line (left scale)
is \(\tau(\sigma)\), the characteristic time scale in years. The orange
line (right scale) is the one year growth rate \(e^{t/\tau}-1\). Again
we see that beyond 75\% vol, things start heating up, and become
outright violent beyond 150\% vol.

\includegraphics[width=12cm,keepaspectratio]{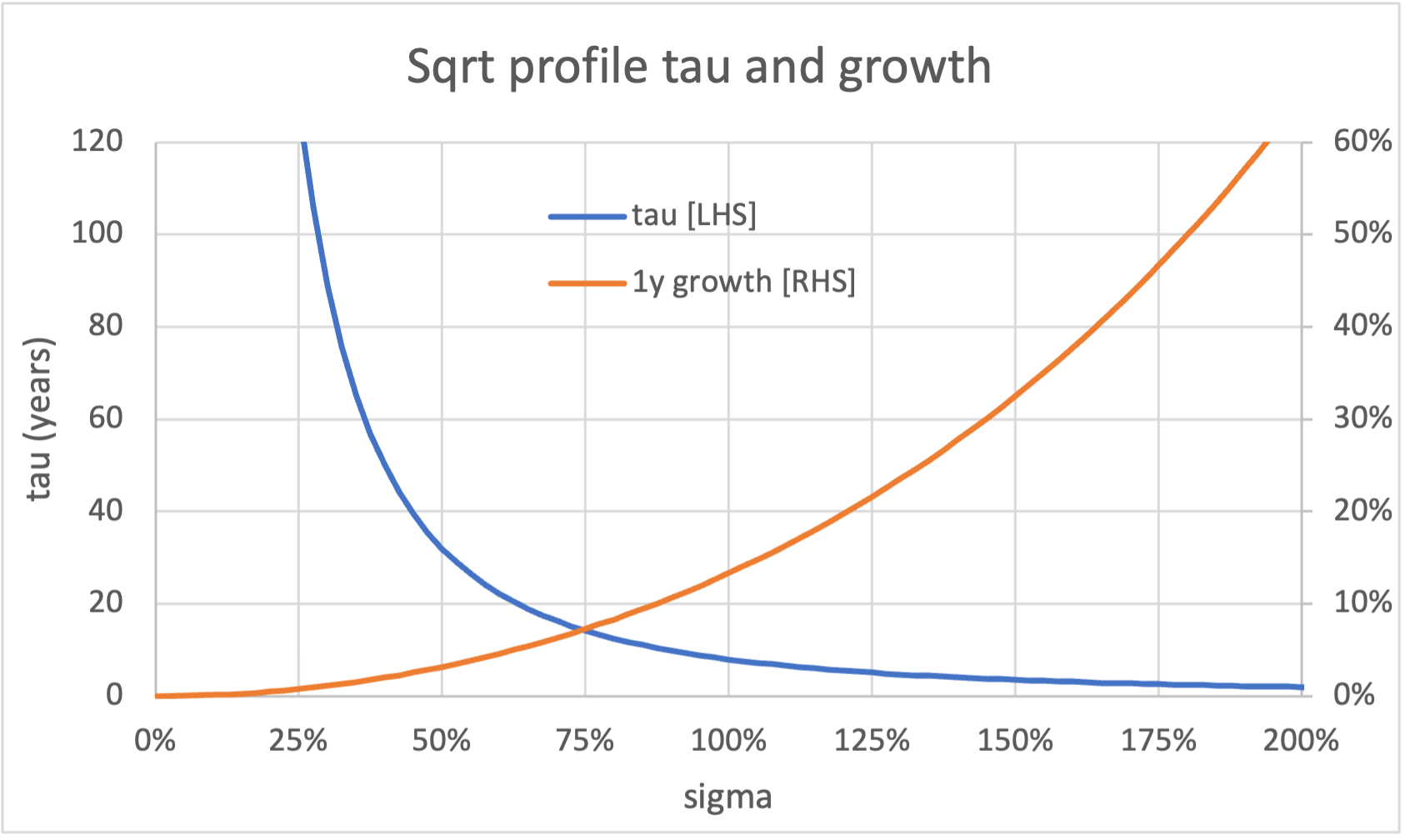}

Finally we are looking at a modified weight AMM. Below we are plotting
the coefficient \(\rho\) from the Black Scholes equation above which
is\\
\[
\rho(\alpha;\sigma) = 
\frac{1}{2}\sigma^2 \alpha (1-\alpha)
\]

and which is the exponential growth rate of the \(\xi^\alpha\) profile
where \(\alpha\) is the weight factor. Note that the equation it is
symmetric around \(\alpha=50\%\) so we cut of the right half. We again
see the strong impact of the volatility, and we also see that the growth
rate -- and therefore the moneys lost to arbitrageurs -- are biggest in
the symmetric case (\(\alpha=50\%\)).

\includegraphics[width=12cm,keepaspectratio]{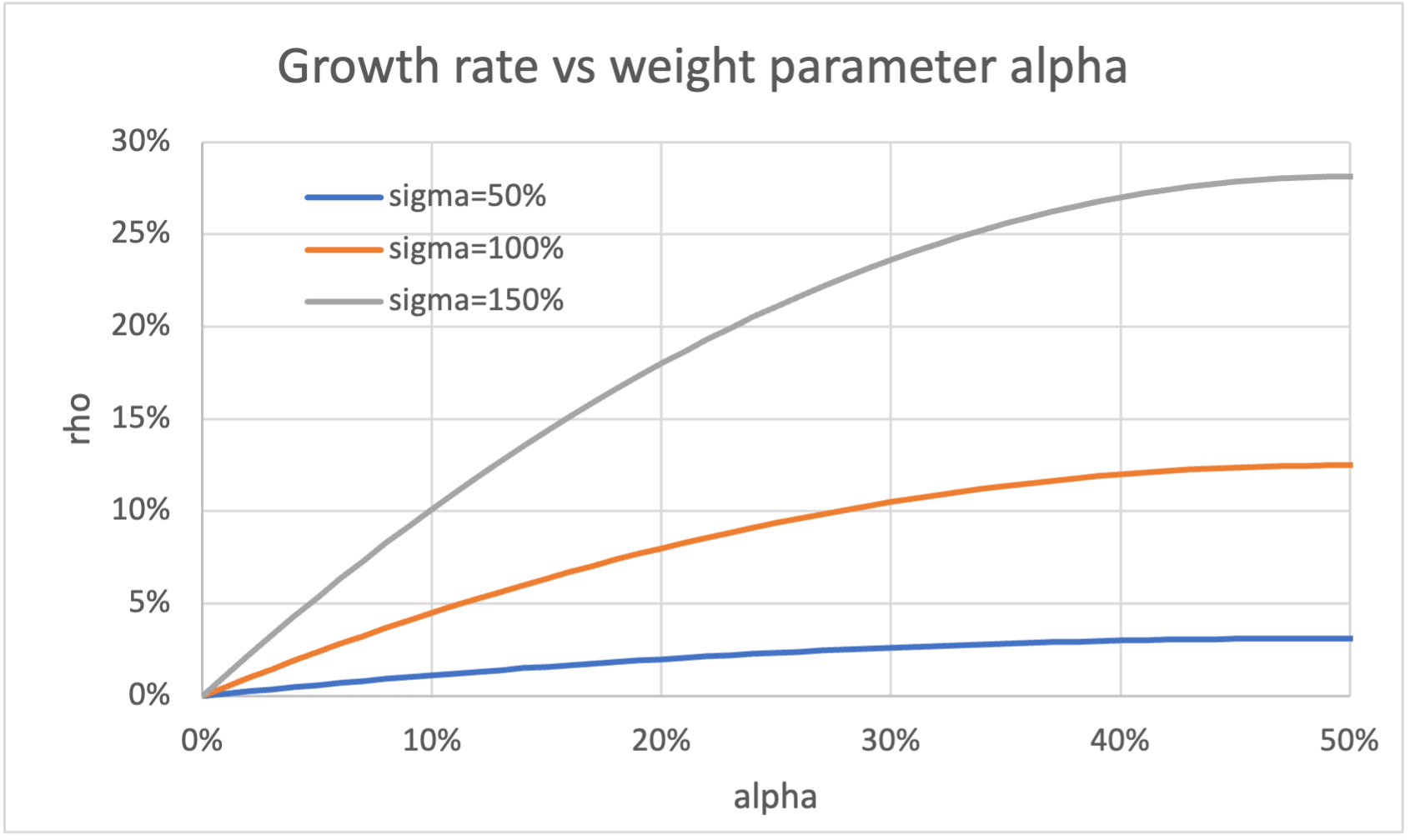}

\hypertarget{mathematics-of-specific-amms}{%
\section{Mathematics of specific
AMMs}\label{mathematics-of-specific-amms}}

\hypertarget{constant-product-kxy}{%
\subsection{Constant product (k=x*y)}\label{constant-product-kxy}}

We now apply the above concepts to the most fundamental of AMMs, the
\textbf{constant product} \(k=x*y\) AMM. As the name implies, this AMM
historically has the \textbf{characteristic function} we already
encountered above, namely

\[
\bar{k} = \bar{f}(x,y) = x*y
\]

where \(x,y\) are the token amounts in their native units respectively.
As before, we consider \(y\) the \emph{numeraire (asset) CSH}, and \(x\)
the \emph{risk asset RSK}. We have seen that there are certain degrees
of freedom in choosing a characteristic function, and we choose an
equivalent one which is

\[
k = f(x,y) = \sqrt{x*y}
\]

This function satisfies the \emph{linearity / homogenity condition}, ie
\(f(\lambda x,\lambda y) = \lambda f(x,y)\). As a consequence, \(k\)
serves as a linear measure of the pool size that is invariant under
trading -- it only changes when liquidity is added (including via fees)
or removed.

We obtain the \textbf{indifference curve} \(y_k(x)\) by isolating \(y\)

\[
y = y_k(x) = \frac{k^2}{x}
\]

and the \textbf{price response function} (\textbf{PRF}) by deriving the
indifference curve with respect to \(x\)

\[
\pi(x) = -\frac{dy}{dx}=\frac{k^2}{x^2}=\frac y x
\]

The price in the PRF is expressed in units of CSH per RSK as one would
expect.

Given a quantity \(x\) of RSK the AMM holds, its holdings in CSH are
\(y=k^2/x\). The value of the RSK holdings, measured in CSH, is
\(x \cdot \pi(x) = x \cdot k^2/x^2=k^2/x\). In other words, the value of
CSH and RSK holdings is the same. This is a key result, which is why we
repeat it:

\begin{quote}
In the constant product AMM in equilibrium with the market, the value of
the CSH holdings always equals that of the RSK holdings
\end{quote}

We denote \(\xi = p/p_0\) the current \textbf{price ratio} of the pool,
ie the ratio of the current price \(p\) divided by \(p_0\), the price at
the time the pool was seeded (we assume a single seeding event for
simplicity; we \emph{can} do that because as we discussed previously, we
can consider every position its own little micro AMM, and the actual AMM
being the combination of those). The \textbf{portfolio value ratio}
\(\nu(\xi)\) is

\[
\nu(\xi) = \sqrt{\xi}
\]

The normalized portfolio value \(\nu\) here is defined as the ratio of
the current portfolio value and the initial of the portfolio value, ie
\(\nu(t=0)=1\).

We have previously proven the above formula, but we'll show it again
here for ease of reference. It is easiest to work our way backwards. It
is well known that there is a correspondance between option profiles and
hedge portfolios. More specifically, in order to hedge a payoff profile,
the replicating strategy holds Delta units of the risk asset. The
remainder of the portfolio value then is held in the numeraire asset.
The Delta is the derivative of the profile with respect to the price. As
this is in units of the risk asset it must be multiplied with \(\xi\) to
be converted into the numeraire. We find that the Cash Delta

\[
\Delta_{\mathrm{cash}} =
\xi \frac {d}{d\xi} \sqrt{\xi} =
\frac 1 2 \sqrt \xi
\]

In other words, half of the value is invested in the risk asset, and
therefore the other half must be invested in the numeraire asset. Going
backwards this means that if we hold at all times the same amount in the
numeraire and the risk asset, our payoff profile will the the square
root profile. This concludes our proof.

From the above we can calculate the normalized \textbf{cash strike
density function} which is \(\xi \nu''(\xi)\), ie

\[
\mu_{\mathbf{cash}}(\xi) = -\frac{1}{4 \sqrt \xi}
\]

and the normalized \textbf{cash Gamma} is \(\xi^2 \nu''(\xi)\), ie

\[
\Gamma_{\mathrm{cash}}(\xi) = -\frac{1}{4} \sqrt \xi
\]

To calculate the \textbf{divergence loss} (commonly but mistakenly
referred to as \textbf{impermanent loss} and defined as the difference
between the HODL value of the initial portfolio and the value of the AMM
porfolio) we note that the initial portfolio was 50:50 in CSH:RSK. The
HODL value of that initial portfolio (ie its value had the portfolio
composition not changed) behaves like \(\frac{1+\xi}{2}\), with our
normalization \(\xi(t=0)=1\). The divergence loss \(\Lambda\) is
therefore

\[
\Lambda(\xi) = \frac{1+\xi}{2}-\sqrt{\xi}
\]

Below we are presenting a few charts. First we draw \(\Lambda(\xi)\)
against \(\xi\) on a very wide scale, of up to 40x changes in price. We
see that the curve quickly becomes linear and unbounded on the upside,
which is unsurprising as the linear term in \(\xi-\sqrt \xi\) dominates.

\includegraphics[width=12cm,keepaspectratio]{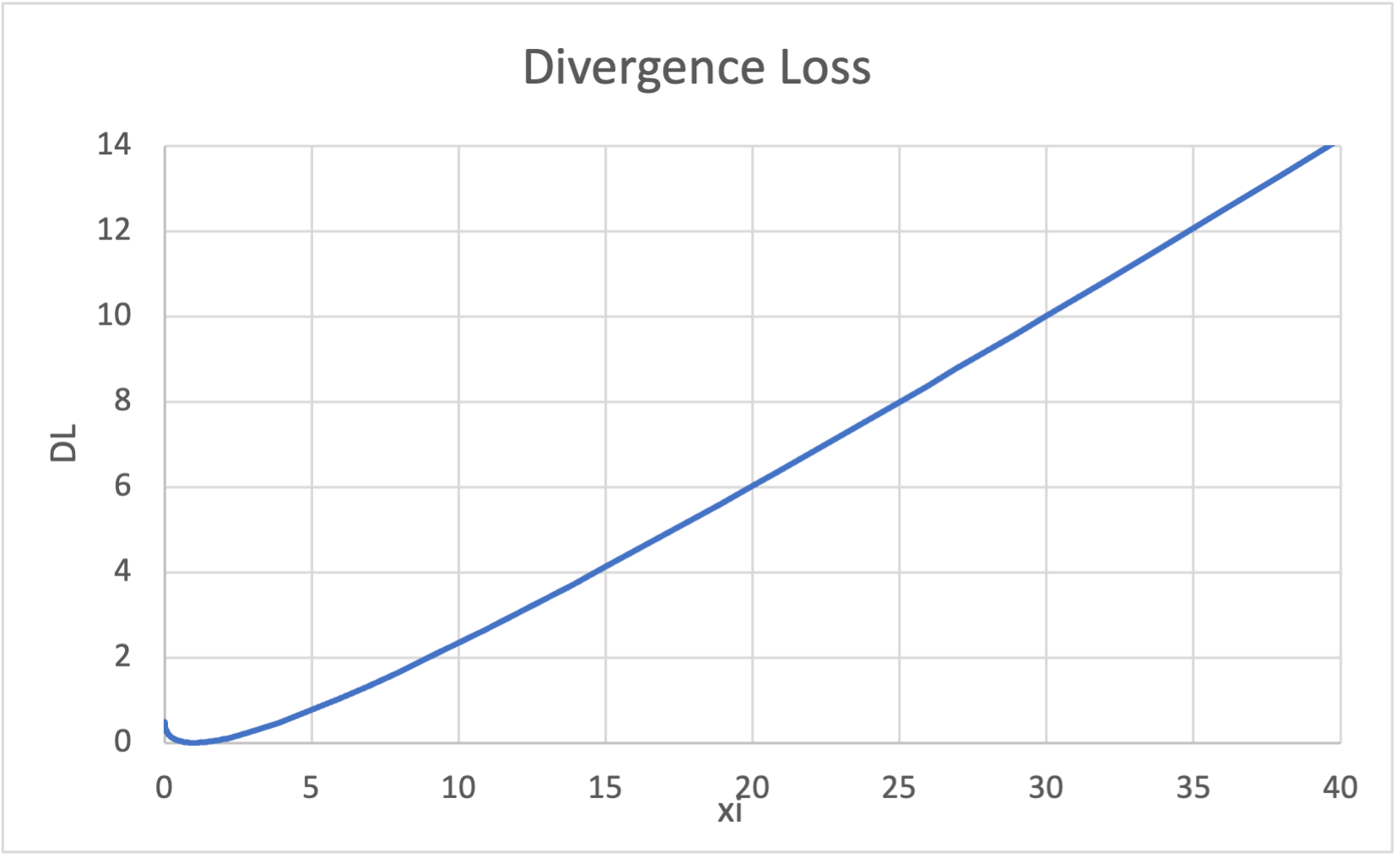}

The picture for more realistic changes in value is more interesting:
\(\Lambda(\xi)\) is relatively flat for changes between minus and plus
50\% (in this range we have \(\Lambda < 5\%\)). \(\Lambda\) only becomes
a significant issue for major divergence. On the downside, \(\Lambda\)
is bounded at 50\%. The reason for this is not the limited loss on the
AMM portfolio -- it goes to zero when \(\xi\) goes to zero. It is rather
because the HODL position loses 50\% (it is initially invested 50\% in
the risk asset), and therefore the maximum possible loss against HODL is
50\%.

\includegraphics[width=12cm,keepaspectratio]{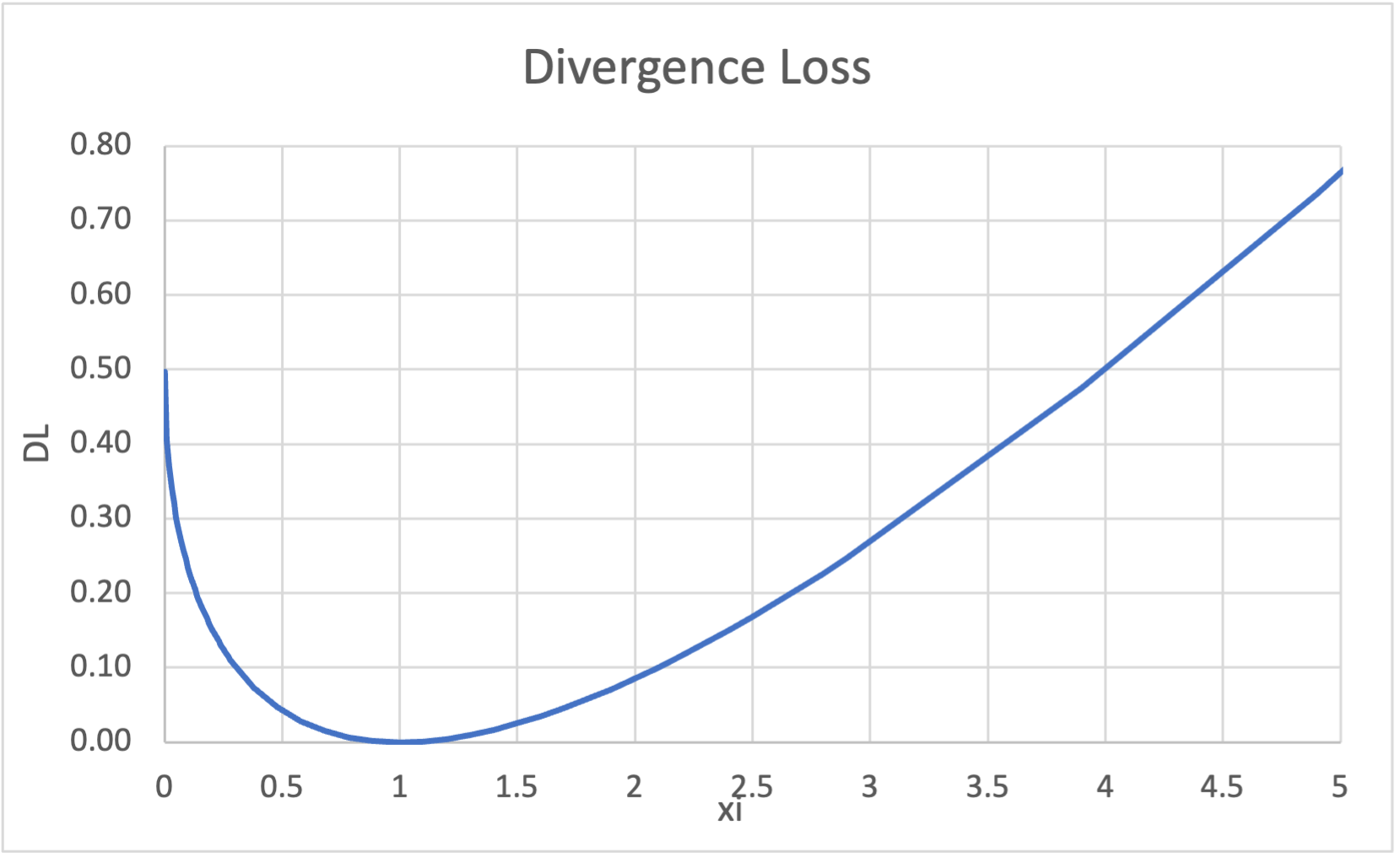}

Note that the asymmetry in the chart is entirely due to the choice of
numeraire. As shown by the characteristic function, the underlying model
is fully symmetric in RSK and CSH, but any choice of numeraire breaks
this symmetry.

In the next chart we show the percentage DL, ie
\(\frac{\mathrm{HODL}-\mathrm{AMM}}{\mathrm{HODL}}\). By construction
this is bounded at unity, and we reach those bounds towards both ends.
However -- on the upside it takes a while to get there, as after 10-15x
the curve flattens considerably.

\includegraphics[width=12cm,keepaspectratio]{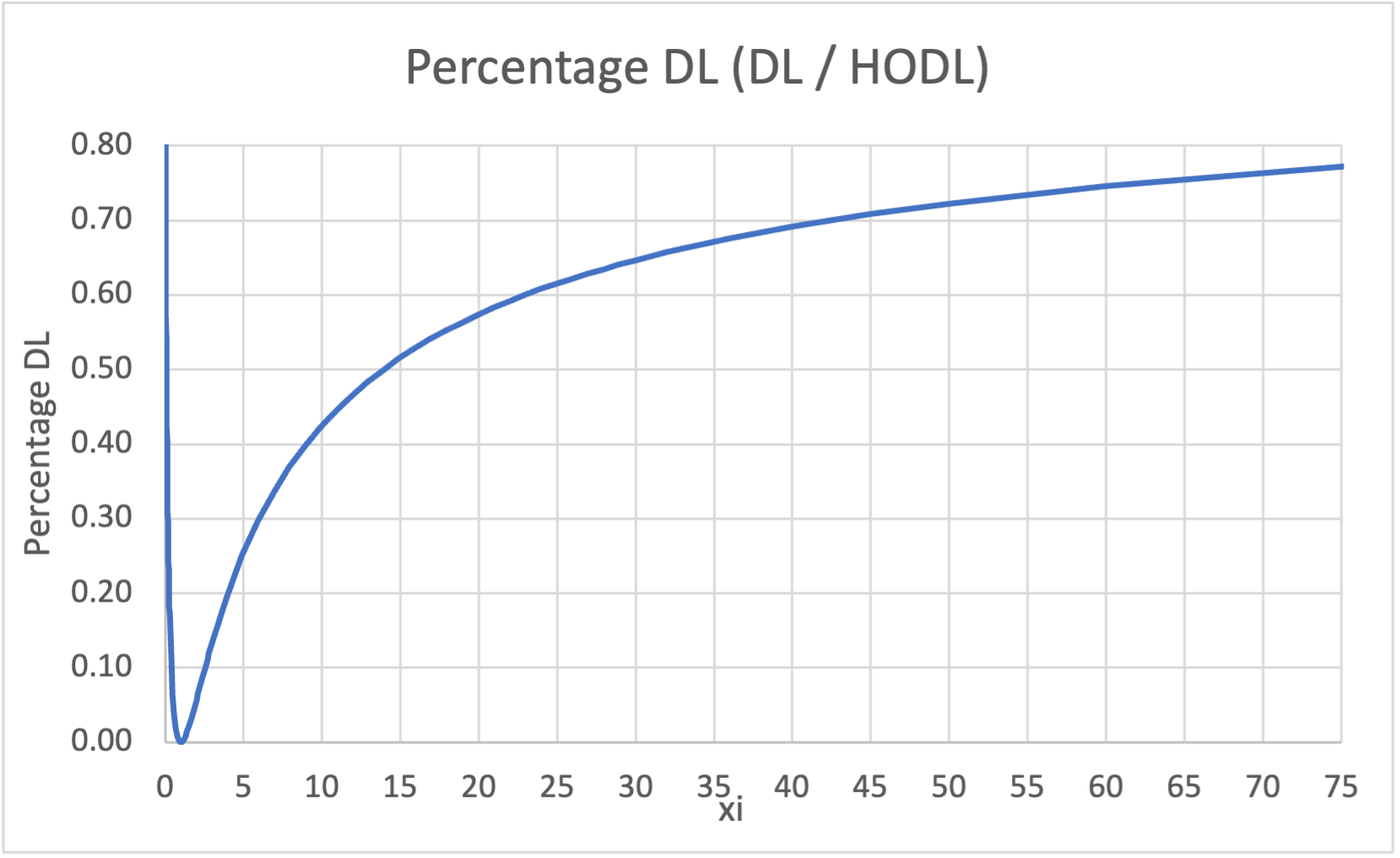}

Below we show the same chart on a smaller scale showing that on the
upside the DL is acceptable for moderate moves, but it becomes painful
for say 4x where it is beyond 20\%

\includegraphics[width=12cm,keepaspectratio]{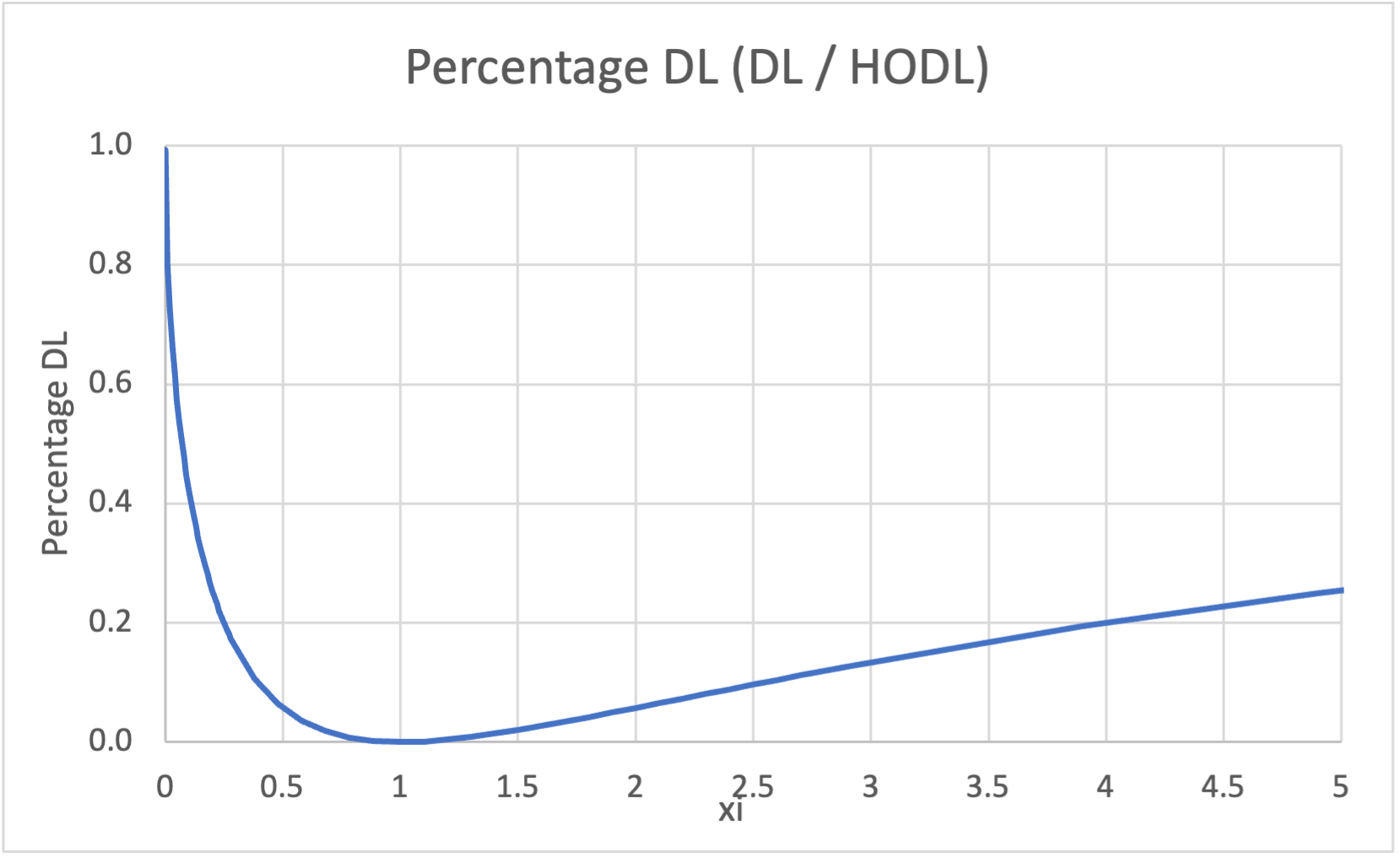}

\hypertarget{constant-sum-kxy}{%
\subsection{Constant sum (k=x+y)}\label{constant-sum-kxy}}

Before moving on to variations of the \emph{constant product} AMM we
want to make a quick detour via the \textbf{constant sum} AMM with the
\textbf{characteristic function}

\[
f_p(x,y) = p*x + y
\]

It is easy to see that this AMM has a linear \textbf{indifference curve}
that crosses the axis at \((0, k)\) and \((k/p, 0)\). The \textbf{price
response function} \(\pi_p(x)\) is the constant function

\[
\pi_p(x) = p
\]

meaning that, ignoring fees, the AMM always buys and sells at the same
price \(p\). Of course the AMM can run out of either CSH or RSK in which
case it will be stuck at the boundary until someone is willing to trade
with it in the right direction. Within the range this AMM has neither
slippage nor Gamma and therefore it does not offer arbitrageurs an
incentive to bring it back into equilibrium. An unmodified constant sum
AMM will generally be stuck at one of its boundaries most of the time.

The \textbf{portfolio value function} of the constant sum AMM is

\[
\nu(\xi) = min(\xi, 1)
\]

This is because it will always be 100\% invested in the underperforming
asset, ie in CSH on the upside, and in RSK on the downside. Note that
the above profile is a short option profile with a strike at \(\xi=1\).

The \textbf{strike density function} is

\[
\mu(\xi) = \delta(\xi-1)
\] where \(\delta\) is the Kronecker delta function, ie a ``function''
(in the physicist sense of the word) that has surface one, is infinity
at \(\xi=0\), and zero everywhere else. Because the strike is at 1, the
\textbf{cash strike density function} is the same as the strike density
function because the \(\xi\) in front of the \(\delta\) can be replaced
with 1. However, we choose to not make this replacement at that time
because this is idiosyncratic to our normalization choices. \[
\mu_{\mathbf{cash}}(\xi) = \xi \delta(\xi-1)
\]

The \textbf{cash Gamma}, where we've made the same choice with respect
to the \(\xi^2\) term, is below

\[
\Gamma_{\mathrm{cash}}(\xi) = \xi^2 \delta(\xi-1)
\]

Finally, note that if we prefer a soft border rather than an AMM that
gets stuck at a boundary, we can choose a modified indifference curve
like for example

\[
f_{p;\varepsilon} = max(p*x+y, \varepsilon * x * y)
\]

that in the boundary regions \(x\to 0\) and \(x \geq k/p\) behaves like
a constant product AMM and therefore never runs out of collateral. In
this case the AMM experiences DL etc just like the constant product AMM
whilst it is in the boundary region. This is very similar to the way the
stableswap algorithm works that we'll discuss below.

\hypertarget{concentrated-range-bound-and-levered-liquidity}{%
\subsection{Concentrated, range-bound and levered
liquidity}\label{concentrated-range-bound-and-levered-liquidity}}

We will now discuss the concepts of concentrated, range-bound and
levered liquidity. Those concepts are closely related by they are not
exactly the same:

\begin{itemize}
\item
  \textbf{Range-bound liquidity.} Range-bound liquidity is liquidity
  that is only available to make markets in a certain price range;
  typically the range is limited at both sides but that does not have to
  be the case.
\item
  \textbf{Levered liquidity.} Levered liquidity is liquidity that is
  amplified, for example in a concentrated liquidity setting where
  excess liquidity is removed from a range bound AMM.
\item
  \textbf{Concentrated liquidity.} Concentrated liquidity is a
  range-bound liquidity where \emph{all} excess collateral (ie
  collateral that can never level the pool) is removed; this means that
  outside the range the AMM only holds one of the two assets
\end{itemize}

\hypertarget{range-bound-liquidity}{%
\subsubsection{Range-bound liquidity}\label{range-bound-liquidity}}

As defined just above, range-bound liquidity is liquidity that is only
available to make markets in a certain range. We have already seen and
extreme example of this type above, when we looked at the constant sum
AMM. Another easy application of this concept is a constant product AMM
where some collateral is removed without adusting the indifference
curve. Instead, when the AMM runs out of collateral it stops trading.
More precisely, when it runs out of one token it will no longer sell it
-- it simply cannot as it can't deliver. It however stands ready to buy
that token, provided the price is right.

As an example we have a standard RSK / CSH constant product AMM, and we
remove some CSH. The AMM sells CSH and buys RSK when the RSK price goes
down. This means that when a certain price \(\xi_0\) is reached, the AMM
runs out of CSH and can therefore no longer buy RSK. If the drop of RSK
continues and the price \(\xi<\xi_0\) is outside the range, the AMM
simply pauses. However, as soon as the price \(\xi>\xi_0\) moves back
into the range, the AMM starts buying CSH again and is back in the game.

In this example the range was implied by the removed liquidity. In
practical applications users arguably prefer to specify a \emph{price}
range where to provide liquidity. We have seen above that the price is
directly linked to the collateral outflow, but there is a catch: this
only works on a specific indifference curve. If the AMM moves curves, eg
because liquidity is added or removed or it earns fees, the range
changes (wider with more liquidity and vice versa). There are two
solutions to this conundrum:

\begin{enumerate}
\def\labelenumi{\arabic{enumi}.}
\item
  Do not pay fees into the pool but keep them separate; this reduces
  collateral efficiency, but it ensures that the AMM remains on the same
  indifference curve
\item
  Adjust the pool constant \(k\) such that the apparent liquidity is
  bigger, and that therefore the AMM will run out of tokens at exactly
  the same price point.
\end{enumerate}

The former solution may be more practical within the high gas cost
environment -- and this is for example the way Uniswap v3 does it -- but
we want to briefly discuss the alternative here, at the example of a
constant product AMM. We recall that \(y=k/x\) and the price
\(\pi=k/x^2\). We know want to keep the price constant and solve for
\(k\). An easy calculation yields

\[
x(k) = \sqrt{\frac k \pi},\ \  y(k) = \sqrt{k \pi}
\]

where \(x(k), y(k)\) is a paramterised boundary curve. This curve is a
straight line going through the origin. We have drawn an example in the
chart below. Here the blue, orange and grey curves are indifference
curves at various values of k, and the yellow line is the cutoff point
at a unity price. So if the AMM wants to remain above unity price it can
only use the parts of the curves that are above the yellow line, and if
it wants to trade below unity price it must remain on the parts of the
curves below the yellow line.

\includegraphics[width=12cm,keepaspectratio]{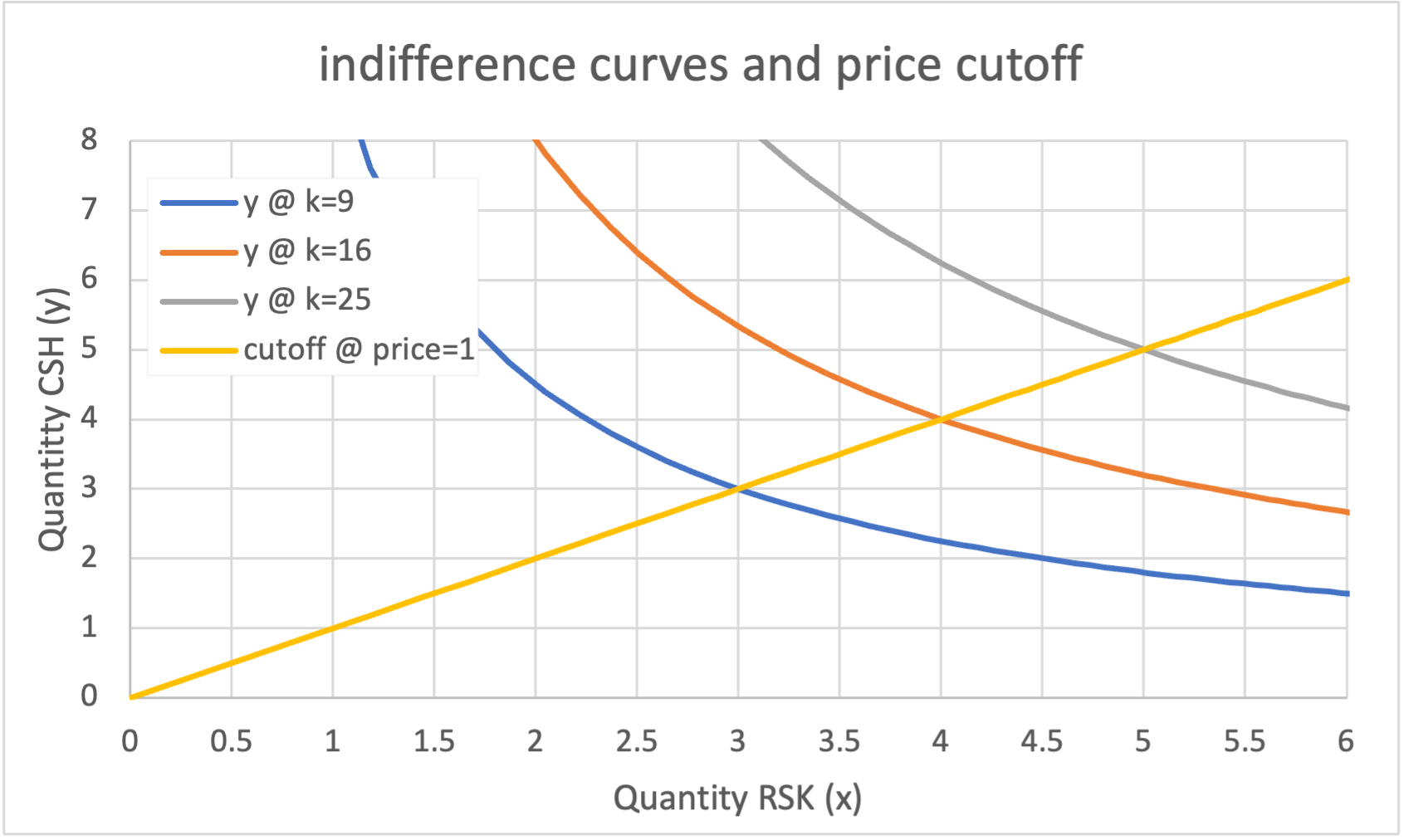}

In practice it may be easier for range-restricted AMMs to operate based
on an \emph{unrestricted} \textbf{characteristic function} and then
explicitly impose the boundary constraints at the level of the
\emph{indifference curve}. However, there may be situations where
operating at the characteristic function level to start with is more
suitable. Below we'll sketch a process for creating a restriced
characteristic function from an unrestricted one and a boundary
condition.

We are looking specifically at a \emph{lower} bound for the amount of
RSK held, and we call this boundary \(x_0\). The intuition behind this
contstruction is that, once we get close (or slightly beyond) \(x_0\),
we are transported on the fast track to \(x=0\). We do this by defining
a function \(\hat x_\varepsilon(x)\) which is described in the chart
below: for \(x>x_0\), ie in the desired range, we have
\(\hat x_\varepsilon(x)=x\). For \(x<x_0\) the turbo kicks in however,
and the function is

\[
x<x_0 \Rightarrow x_\varepsilon(x) = x_0 - \frac{x-x_0}\varepsilon
\]

For \(\varepsilon=1\) we simply have \(\hat x_1(x)=x\). The magic
happens for \(\varepsilon \rightarrow 0\): the smaller \(\varepsilon\),
the faster \(\hat x_\varepsilon(x)\) is at zero.

\includegraphics[width=12cm,keepaspectratio]{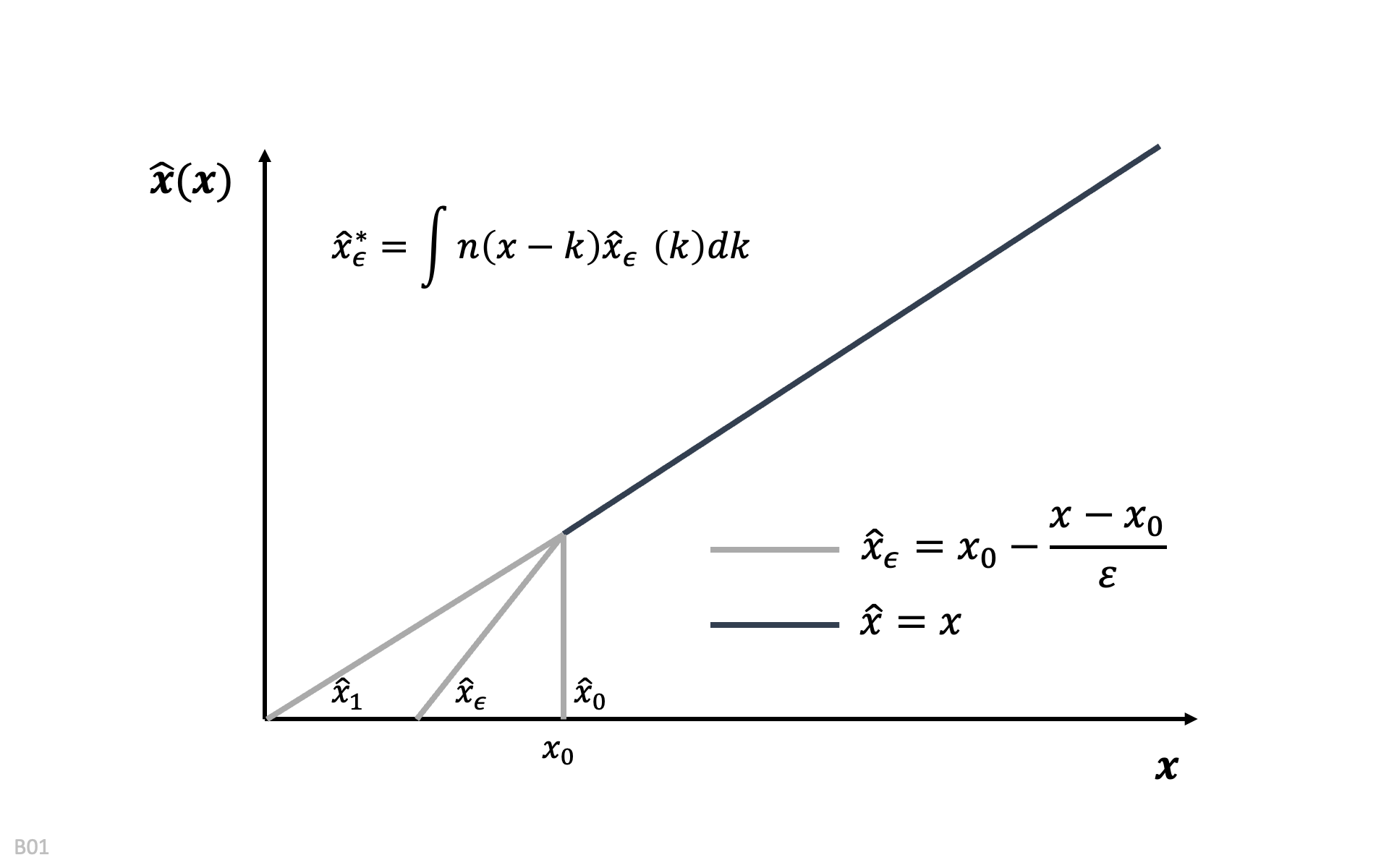}

So instead of using \(f(x,y)=k\) we use

\[
f(\hat x^*_\varepsilon(x), \hat y^*_\varepsilon(y)) = k
\]

where the \(x^*_\varepsilon(x), y^*_\varepsilon(y)\) are defining a
suitable range, and \(\varepsilon\) is very small but not zero. The
asterisk denotes that additionally we are using a convolution
\(\int n(x-k) \hat x^*_\varepsilon(k) dk\) with a suitable Gaussian
kernel \(n(x)\) to make the function \(C^\infty\). This may be overkill
however, and the initial \(C^0\) solution \(\hat x_\varepsilon(x)\) may
just work fine.

\hypertarget{concentrated-and-levered-liquidity}{%
\subsubsection{Concentrated and levered
liquidity}\label{concentrated-and-levered-liquidity}}

Once we have restricted the range, and therefore limited the outflow of
one or both tokens from the restricted AMM, we can apply leverage by
simply removing those tokens, or never even contributing them in the
first place. Whilst inside the range, the AMM will behave as if the
liquidity had not been removed. So everything else being equal, the fees
earned per unit of liquidity contributed will be higher, but so will be
the divergence loss. Ultimately Miller-Modigliani applies (figuratively,
not actually) in the sense that \emph{leverage does not economically
matter in efficient markets}, and whilst this leverage increase the
returns it also increases the risk, so the risk-adjusted returns remain
the same.

We are now looking at details for the levered constant product AMM, ie
the Uniswap v3 model. As before, \(\xi\) is our normalized price ratio,
and \(\xi_0, \xi_1\) is the liquidity range, using the same
normalization. To understand this intuitively, \(\xi\) will start at
100\%, and \(\xi_0, \xi_1 = 80\%, 120\%\) means a range that is 20\% up
and down from the starting price. We have also added a portfolio
notional factor \(n_0\) here that we can use to normalize our portfolio
value \(\nu\) however we want. The AMM portfolio value \(\nu\) below /
in / above the range respectively is given by the three equations below
(see eg {[}Lambert21{]})

\[
\nu(\xi) = n_0 \cdot \xi \ \ \ \forall \xi < \xi_0
\]

\[
\nu(\xi) = n_0 \cdot \left( \sqrt{\xi_0 \xi_1} \cdot \frac{\sqrt{\xi}-\sqrt{\xi_0}}{\sqrt{\xi_1}-\sqrt{\xi_0}}
+ \sqrt{\xi_0 \xi} \cdot \frac{\sqrt{\xi_1}-\sqrt{\xi}}{\sqrt{\xi_1}-\sqrt{\xi_0}} \right)
\ \ \ \forall \xi_0 < \xi < \xi_1
\]

\[
\nu(\xi) = n_0 \cdot \sqrt{\xi_0 \xi_1} \ \ \ \forall \xi > \xi_1
\]

Note that in the limit \(\xi_0 \rightarrow 0\),
\(\xi_1 \rightarrow \infty\) we find \(\nu(\xi) = \sqrt{\xi}\), provided
we set \(n_0 = (\sqrt{\xi_1}-\sqrt{\xi_0})/\sqrt{\xi_0 \xi_1}\) to
account for the different normalization in this formula compared to the
standard one.

It is interesting to look at the portfolio composition. The holdings of
\emph{risk asset} RSK corresponding to the aforementioned valuation
formula are \[
N_r(\xi) =  n_0 \cdot 1 \ \ \ \forall \xi < \xi_0
\]

\[
N_r(\xi) = n_0 \cdot \sqrt{\frac{\xi_0}{\xi}} \cdot \frac{\sqrt{\xi_1}-\sqrt{\xi}}{\sqrt{\xi_1}-\sqrt{\xi_0}}
\ \ \ \forall \xi_0 < \xi < \xi_1
\]

\[
N_r(\xi)  = 0 \ \ \ \forall \xi > \xi_1
\]

and we see that the normalization of those formulas is such that for
\(n_0=1\), we hold exactly 1 unit of RSK below the range. This is why we
cannot simply set \(\xi_0=0\) in the formula above but we need to adjust
\(n_0\) to have the correct normalization. Above the range we hold no
RSK but only CSH.

For the \emph{numeraire asset} CSH we find

\[
N_n(\xi) =  0 \ \ \ \forall \xi < \xi_0
\]

\[
N_n(\xi) = n_0 \cdot \sqrt{\xi_0 \xi_1} \cdot \frac{\sqrt{\xi}-\sqrt{\xi_0}}{\sqrt{\xi_1}-\sqrt{\xi_0}}
\ \ \ \forall \xi_0 < \xi < \xi_1
\]

\[
N_n(\xi) = n_0 \cdot \sqrt{\xi_0 \xi_1} \ \ \ \forall \xi > \xi_1
\]

so below the range we hold no CSH, and above we hold everything in CSH,
converted at a rate of \(\sqrt{\xi_0 \xi_1}\) which is the effective
conversion price (geometric average of the range boundaries) when moving
through the range.

Below we have graphed the AMM portfolio value \(\nu(\xi)\) against
\(\xi\) for a number of different ranges. Those ranges are symmetric, in
a geometric average sense, around \(\xi=100%
\). Starting with the yellow curve, the widest of the ranges, we see
that we get close to the square root profile. We however start with a
finite slope at \(\xi=0\) where the square root profile starts
vertically.

Asymptotically all profiles start linear (100\% investment in the risk
asset on the risk asset downside) and end up flat (100\% investment in
the numeraire on the upside; they all end up at the same level). Within
the range there is an arc connecting the two asymptotics and
unsurprisingly, the wider the range the further this arc deviates from
the asymptotics.

\includegraphics[width=12cm,keepaspectratio]{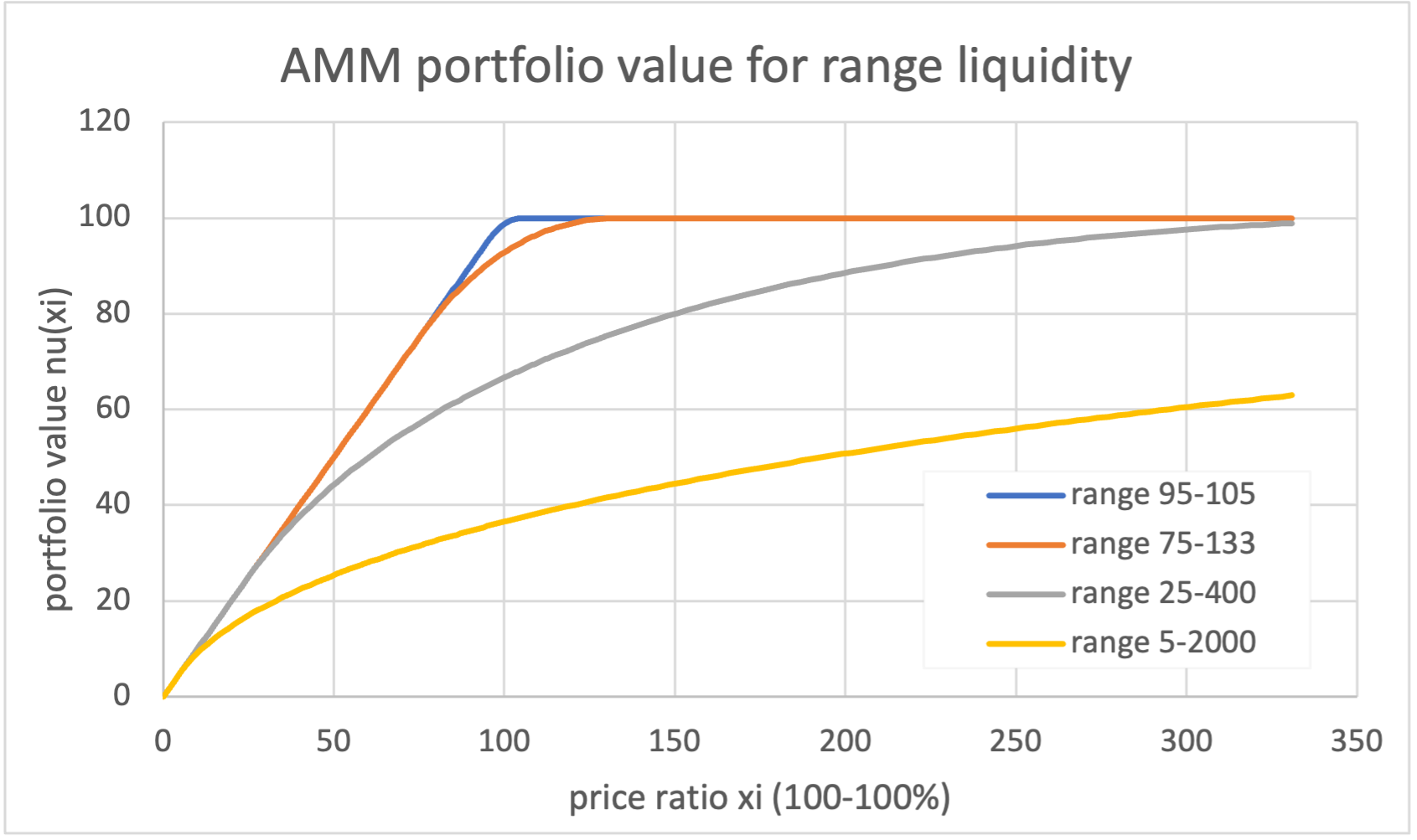}

\hypertarget{divergence-loss-with-concentrated-liquidity}{%
\subsubsection{Divergence loss with concentrated
liquidity}\label{divergence-loss-with-concentrated-liquidity}}

Divergence loss within a concentrated, leveraged liquidity is somewhat
complex. This is not so much for technical reasons, but for reasons of
interpretation. In the frameworks we have discussed so far, and those
that we will discuss below, the liquidity pool composition was fixed at
either 1:1 for the constant product pool, or at \(\alpha:1-\alpha\) for
the modified weights pools. A concentrated pool however by design has a
time-varying pool composition -- 100\% in one of the assets outside the
range, and shifting from one asset to the other within.

The issue with this is that it is not entirely clear what the reference
portfolio against which the DL is computed should be. Another problem
that already arises in other scenarios, and that comes back here on
steroids is how to calculate DL for (a) positions withdrawn in the past,
and (b) a position that have been adjusted during their lifetime. Also
we need to decide (c) whether we only account for DL within the range
when the position earns fees, or also outside of it.

A good candidate for DL computation is \emph{use whatever the portfolio
was when the position was created} (and, as a corrolary: \emph{whenever
the position was adjusted, treat it as if it was withdrawn and
recreated}). Let's do some calculations here and go through that step by
step

\begin{enumerate}
\def\labelenumi{\arabic{enumi}.}
\item
  The position is initialised at 1 RSK = 150 CSH at inception, and the
  range is from 95 to about 105, with the geomtric center at exactly
  100; the portfolio is 100\% in CSH, holding 150 CSH
\item
  The price drops to just above the range; the portfolio is still the
  original portfolio (and in any case, 100\% CSH), therefore no DL
\item
  The price drops to just below 95; now the portfolio is 100\% in RSK,
  exchanged at a price of 100; in other words it holds 1.5 RSK which, at
  a price of 95, are worth 142.5 CSH. This is a DL of 7.5 CSH.
\item
  The price drops to 50. The portfolio does not change (and does not
  earn fees) so it still holds 1.5 RSK, worth 75 CSH. The DL is now 75
  CSH.
\item
  Scenario A: the position is unwound; Scenario B: the position remains.
\item
  RSK recovers to its original value of 150 CSH. In Scenario A, DL
  disappears and is exactly zero. Scenario B is complex:

  \begin{enumerate}
  \def\labelenumii{\arabic{enumii}.}
  \tightlist
  \item
    If the DL was crystallized CSH at 75 it remains at 75 CSH
  \item
    If the DL was crystallized in CSH at 7.5 (because beyond that the
    positions was de facto inactive) it remains at 7.5 CSH
  \item
    If the losses are translated into RSK at the then prevailing
    exchange ratio, which is 1.5 RSK or 0.15 RSK respectively, and then
    crystallized, the DL is 225 or 22.5 CSH respectively
  \item
    If the position is carried forward as 1.5 RSK the DL becomes a gain
    of 0.5 RSK, or 75 CSH
  \end{enumerate}
\end{enumerate}

The distinction between in-range and out-of-range DL depens on the
purpose of the calculation. Once the range has been crossed the position
no longer earns fees, which distorts measures like a fee / DL ratio --
for those \emph{optimal} behaviour of the LPs may be a good working
assumption. To compute actual LP returns however, DL outside the range
should probably be taken into account.

The other choices however all correspond to genuine alternative trading
strategies, and with the DL being an opportunity loss, all of them are
in theory acceptable. There is one caveat however which is that the
Divergence \emph{Loss} should not really become a gain, so if the last
of those definitions is used the measure should probably be renamed.

In our view, the most useful DL measures are the one that crystallyze
either into CSH, or into a joint numeraire when running a multi-pool
analysis which is the one we used in {[}Loesch21{]}. Whether to use
in-range for full IL numbers is a question of judgement; in our view the
in-range IL number is the one better suited for theoretical fee/IL
ratios, and the full number is better for estimating at actual
opportunity losses incurred.

\hypertarget{modified-weights}{%
\subsection{Modified weights}\label{modified-weights}}

The modified weights AMM is very similar to the constant product AMM,
with the crucial difference that now the two assets may have different
weights, which as we will see leads to a different portfolio
composition. The \textbf{characteristic function} of the modified
weights AMM is

\[
k = f(x,y) = x^\alpha*y^{1-\alpha}
\]

with some parameter \(\alpha \in (0,1)\). It is easy to see that we find
the constant product AMM when we set \(\alpha=1/2\), and also that the
characteristic function above scales linearly.

We now define \(\eta(\alpha)\) which helps us to simplify some of the
formulas that will follow.

\[
\eta = \frac{\alpha}{1-\alpha}\ \Rightarrow\ 
\alpha = \frac{\eta}{1+\eta},\ 
\frac{1}{1-\alpha} = 1+\eta
\]

The two charts that follow show the relationship between \(\eta\) and
\(\alpha\).

\includegraphics[width=6cm,keepaspectratio]{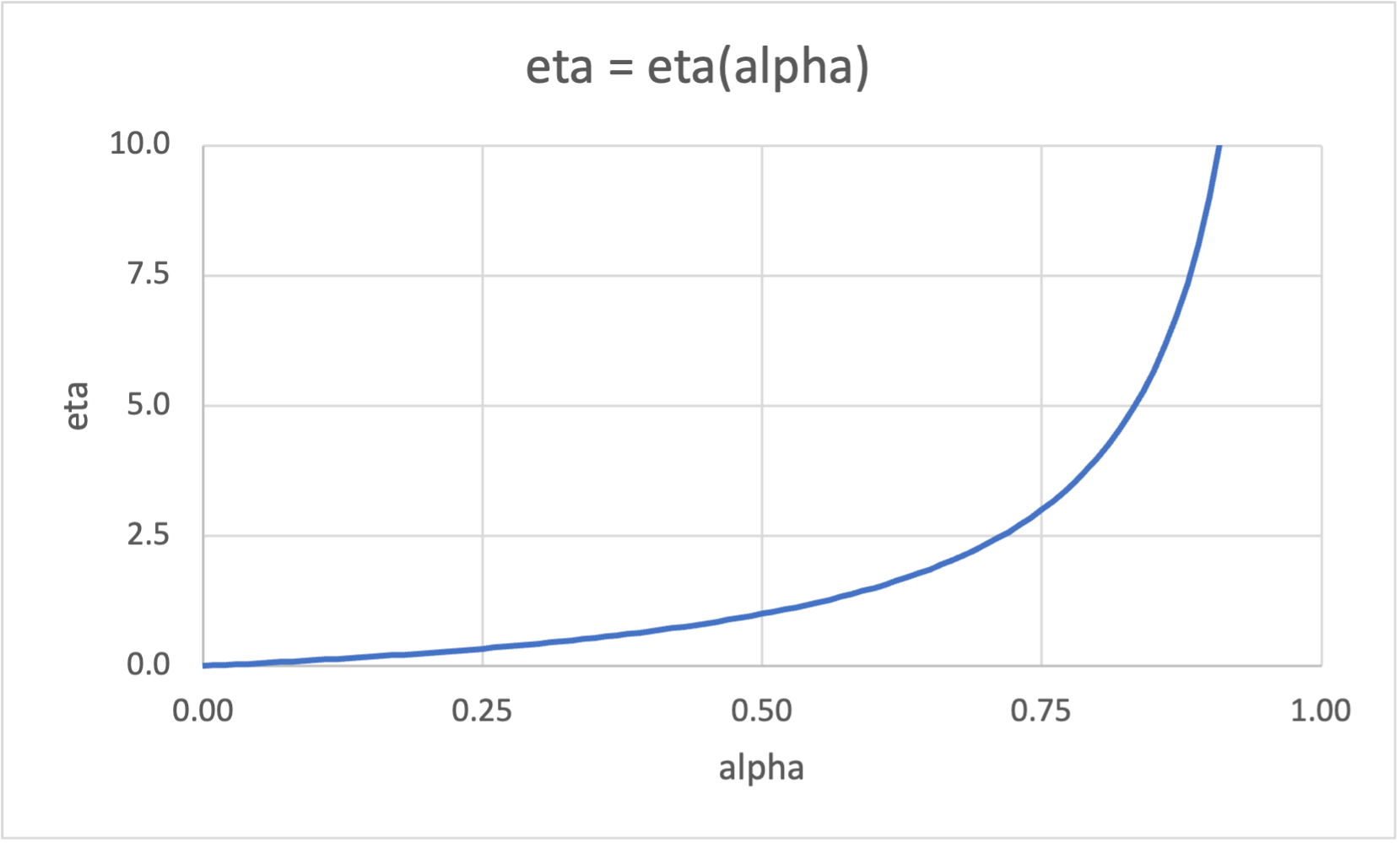}
\includegraphics[width=6cm,keepaspectratio]{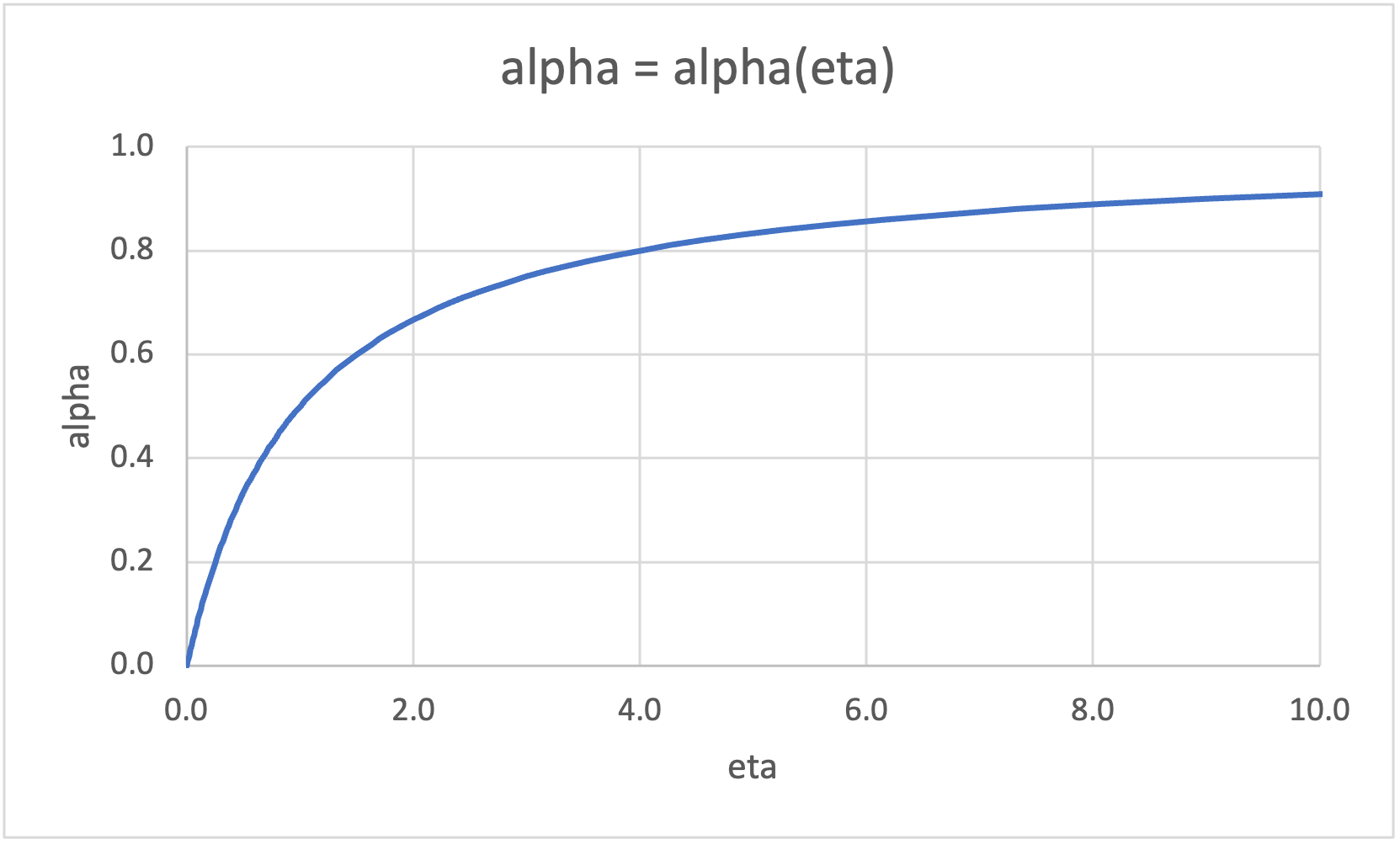}

Using our newly defined \(\eta\) we can write the \textbf{indifference
curve} of the modified weight AMM as follows

\[
y_{k,\alpha}(x) = \left(  \frac k {x^\alpha}   \right)^\frac 1 {1-\alpha}
\]

Alternatively,

\[
y_{k,\eta}(x) = \frac{k^{1+\eta}}{x^\eta}
\]

The \textbf{price response function} in the modified weight case becomes

\[
\pi_{k,\alpha}(x) = 
        \frac{\alpha}{1-\alpha} \left(  \frac k x   \right)^\frac 1 {1-\alpha} =
        \frac{\alpha}{1-\alpha}\   \frac y x   
\]

Alternatively,

\[
\pi_{k,\eta}(x) = 
        \eta \left(  \frac k x   \right)^{\eta+1}=
        \eta\ \frac y x
\]

It is easy to verify that in this case the ratio of the value of the
risk asset and the numeraire asset in the pool is
\(\eta=\frac \alpha {1-\alpha}\). So if \(\alpha>\frac 1 2\) and
therefore \(\eta>1\) we have more risk asset in the pool than numeraire
asset and vice versa.

The portfolio value \(\nu\) in this case becomes \[
\nu = \xi^\alpha = \xi^{\frac \eta {1+\eta}}
\] The proof is mostly the same as the one in the unweighted constant
product case, except that the calculation of Delta (\(=\nu'\)) shows
that the proportion of the portfolio in the risk asset is \(\alpha\) and
therefore \(1-\alpha\) in the numeraire. In other words, the ratio
between risk asset and numeraire is \(\eta\) as it should be.

The \textbf{cash strike density function} is easily calcluated as

\[
\mu_{\mathbf{cash}}(\xi) = 
-\frac{\alpha (1-\alpha)}{\xi^{1-\alpha}}
\]

and \textbf{Cash Gamma} is

\[
\Gamma_{\mathbf{cash}}(\xi) = 
- \alpha (1-\alpha) \xi^\alpha
\]

The \textbf{divergence loss} in the modified weight case becomes

\[
\Lambda_\alpha(\xi) = 1-\alpha + \alpha \xi - \xi^\alpha
\]

and we find our well known formula for the constant product AMM if we
set \(\alpha=\frac 1 2\).

Below we have plotted the DL for different values of \(\alpha\): the
blue line is for the constant product AMM with \(\alpha=\frac 1 2\), the
grey line is for \(\alpha=0.9\) (risk asset to numeraire is 9:1, ie risk
asset prevails), and the orange line is \(\alpha=0.1\) (risk asset to
numeraire is 1:9, ie numeraire asset prevails). When looking at this
analysis we need to remind ourselves however that, even though the
situation is now somewhat asymmetric to start with, our choice of
numeraire does introduce additional asymmetries.

The first chart shows the DL with the risk asset appreciating up to 5x
against the numeraire (from this point onwards the curves continue
mostly linearly to the right). We see that the \(\alpha=0.5\) (constant
product) curve shows a significantly higher DL than the two other
curves. This makes sense as on the one hand the \(\alpha=0.9\) (risk
asset dominant) portfolio predominently holds the risk asset and
therefore does not lag that much if it rallies. The \(\alpha=0.1\)
(numeraire dominant) curve on the other hand measure the DL against a
porfolio that has very little of the risk asset to start with, so again
the relative losses are less.

One the risk asset downside, the terminal value is entirely driven by
the HODL portfolio: the AMM portfolio goes to zero in all three cases.
Therefore the \emph{more} the HODL portfolio loses the \emph{lower} the
DL.

\includegraphics[width=12cm,keepaspectratio]{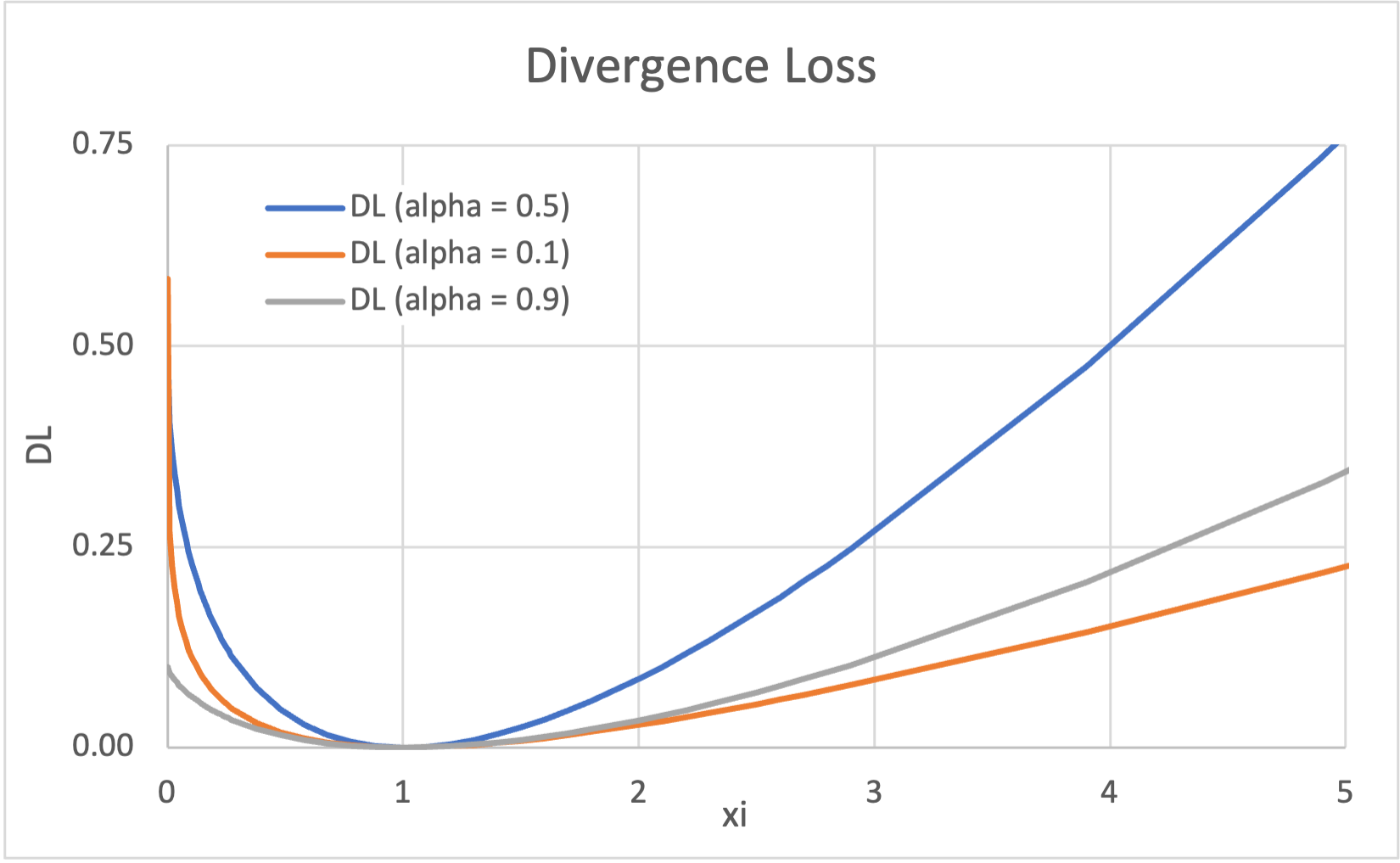}

In the next chart we show the \emph{percentage DL}, ie that DL relative
to HODL, defined as
\(\frac{\mathrm{HODL}-\mathrm{AMM}}{\mathrm{HODL}}\). By construction,
this number must be less or equal 1.0 aka 100\%. This chart shows a very
wide range on the upside, with up to 60x appreciation of the risk asset
versus the numeraire. Again the constant product AMM shows the biggest
losses initially, but the numeraire-dominant AMM (\(\alpha=0.1\))
catches up and even exceeds it eventually. The risk-asset-dominant AMM
(\(\alpha=0.9\)) consistently shows significantly a lower lower
percentage DL than the two others.

\includegraphics[width=12cm,keepaspectratio]{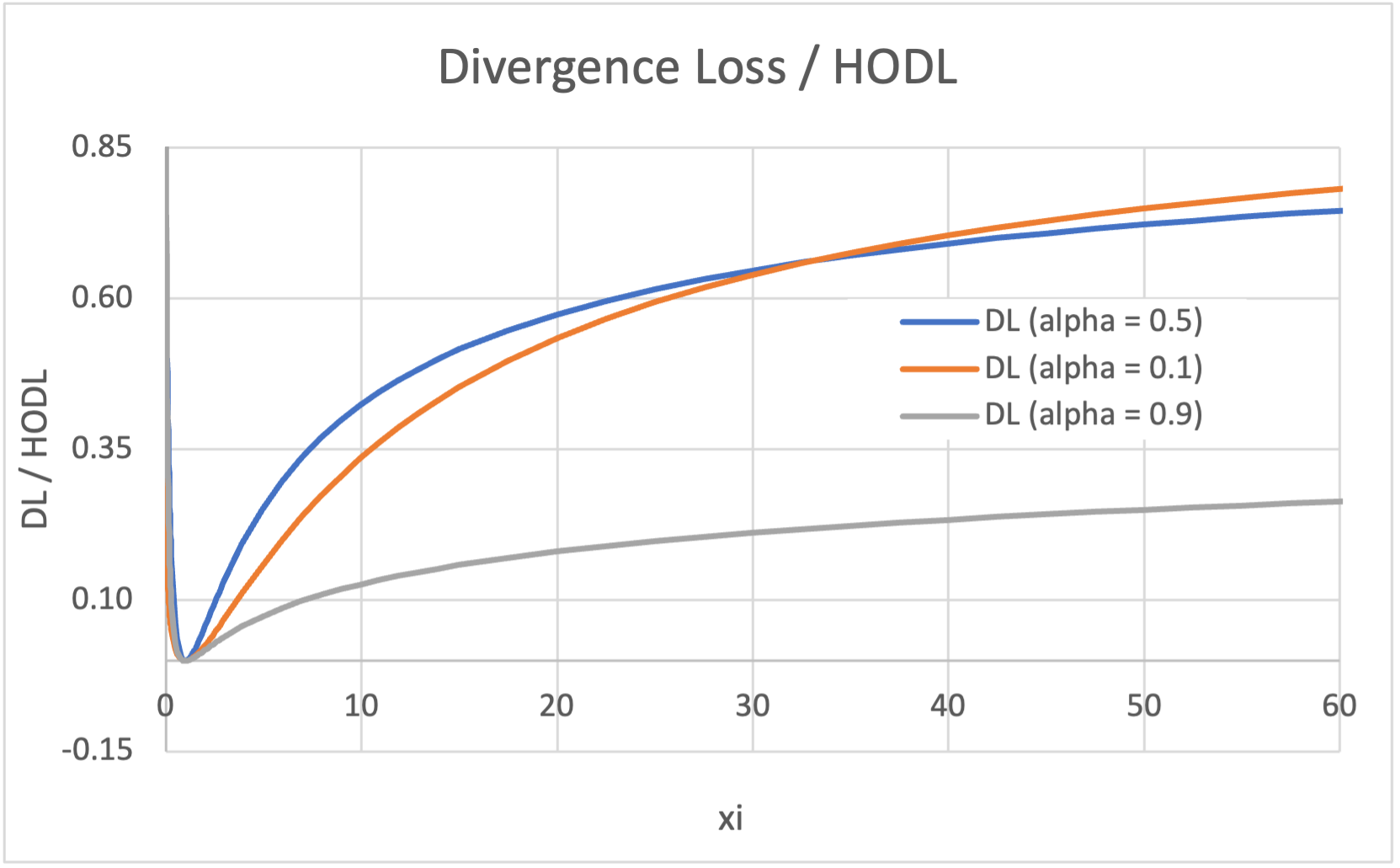}

The final chart here is the same chart as the previous one, but we are
zooming into \(\xi \in (0,1)\) to look at the risk asset downside. We
see that again the constant product AMM performs worst in
\emph{percentage DL}, even though ultimately the risk-asset dominant AMM
(\(\alpha=0.9\)) catches up or even slightly exceeds the losses in
relative terms. The numeraire-dominant AMM (\(\alpha=0.1\)) does
consistently better here.

\includegraphics[width=12cm,keepaspectratio]{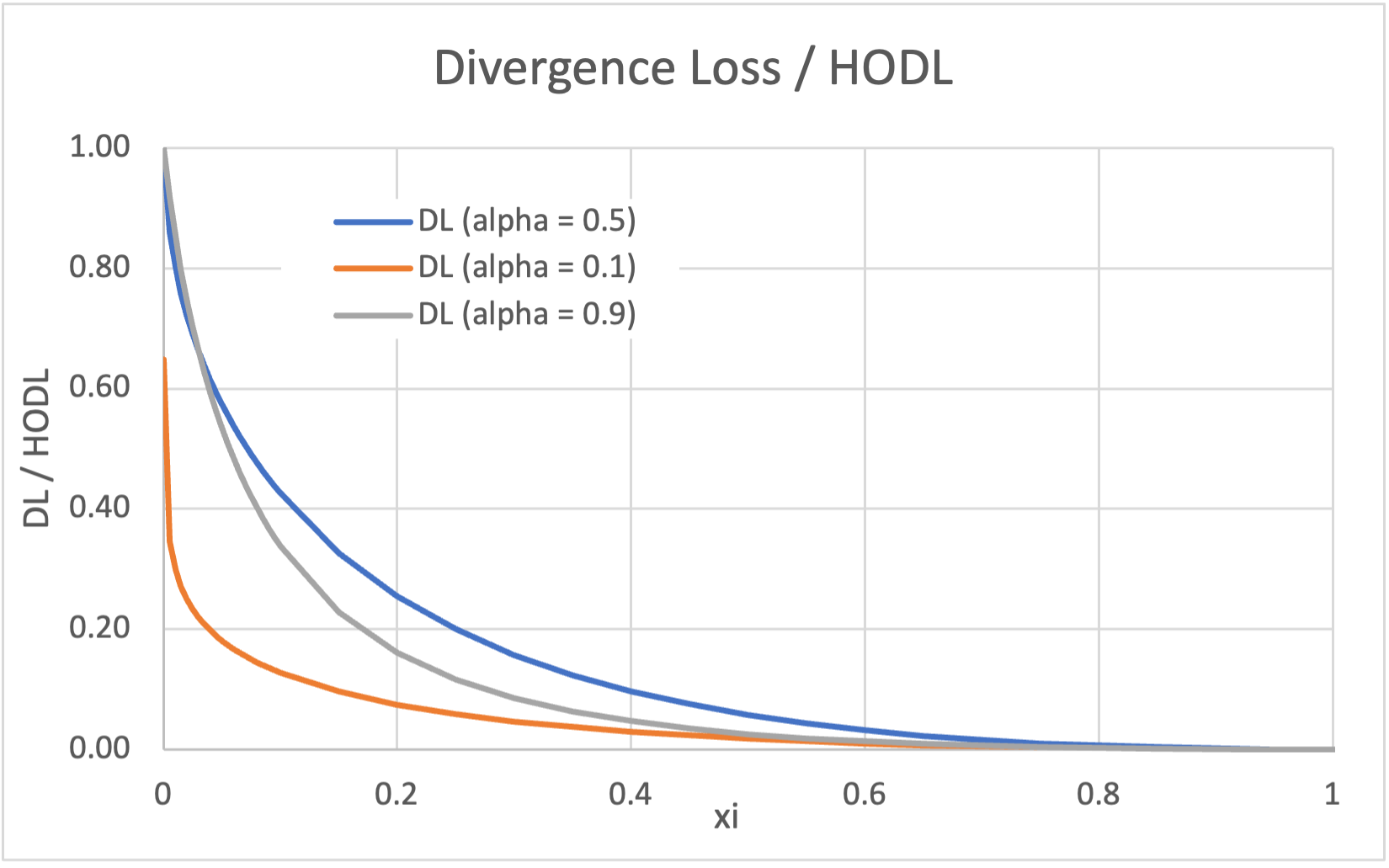}

\hypertarget{modified-curve-aka-2-token-stableswap}{%
\subsection{Modified curve aka (2 token)
stableswap}\label{modified-curve-aka-2-token-stableswap}}

When discussing the constant sum (\(k=x+y\)) AMM which provides
liquidity at one specific price only we already alluded to two different
options what could happen at the boundary

\begin{itemize}
\item
  \textbf{Option 1}: no special treatment at the boundaries
  \(x=0, y=0\), the AMM simply stops trading; that is also the model
  that Uniswap v3 is running at the boundaries of the range
\item
  \textbf{Option 2}: the curve is modified towards the boundary
  \(x\to 0, y\to 0\) such that the characteristic functions always
  become tangent to the axis' and therefore the AMM never runs out of
  assets
\end{itemize}

The stableswap model, introduced by {[}Egorov19{]}, chooses Option 2,
albeit in a multi-token environment. We here go with the reduced
two-token formula described in {[}Niemerg20{]}. In this case, the
\textbf{characteristic function} is

\[
f_{\chi;k} = \chi k (x+y)+xy
\]

where \(\chi\) is a mixing parameter: for \(\chi=0\) the AMM is a
constant product AMM, and in the limit \(\chi\to\infty\) it becomes a
constant sum AMM. Note that the \(k\) is now part of the characteristic
function -- as we'll see in a moment it is there for dimensional reasons
(note that the above characteristic function does \emph{not} have our
usual scaling properties; it scales quadratically).

The \textbf{indifference curves} are now defined by the equation

\[
f_{\chi;k}(x,y) = \chi k (x+y)+xy =k^*(\chi, k) = \chi k^2 + \frac{k^2}4
\]

where for the time being we consider \(\chi\) as previously as a model
parameter. As before, for \(\chi=0\) we recover the constant product
formula, for \(\chi \to \infty\) the constant sum, and the \(x,y,k\) all
scale linearly under the scale symmetry we previously discussed, as per
the design goals in {[}Egorov19{]}.

The key difference here when compared to the other AMMs we have
discussed so far is that \(k\) is now part of the characteristic
function and therefore also part if the indifference curve, ie it
appears on the right hand side of the equation. For a single
indifference curve (\(k\) fixed) this does not matter. However, if \(k\)
changes, eg when assets are contributed to or withdrawn from the pool,
then the characteristic function changes as well. Therefore a pair
\(x,y\) no longer necessarily determines the state of the AMM -- we may
need to specify \(k\) in addition to \(x,y\) as there may be multiple,
or even an infinite number of \(k\) that lead to the same portfolio
composition.

This chart from {[}Egorov19{]} shows the shape of the above curve
compared to constant product and constant sum:

\includegraphics[width=12cm,keepaspectratio]{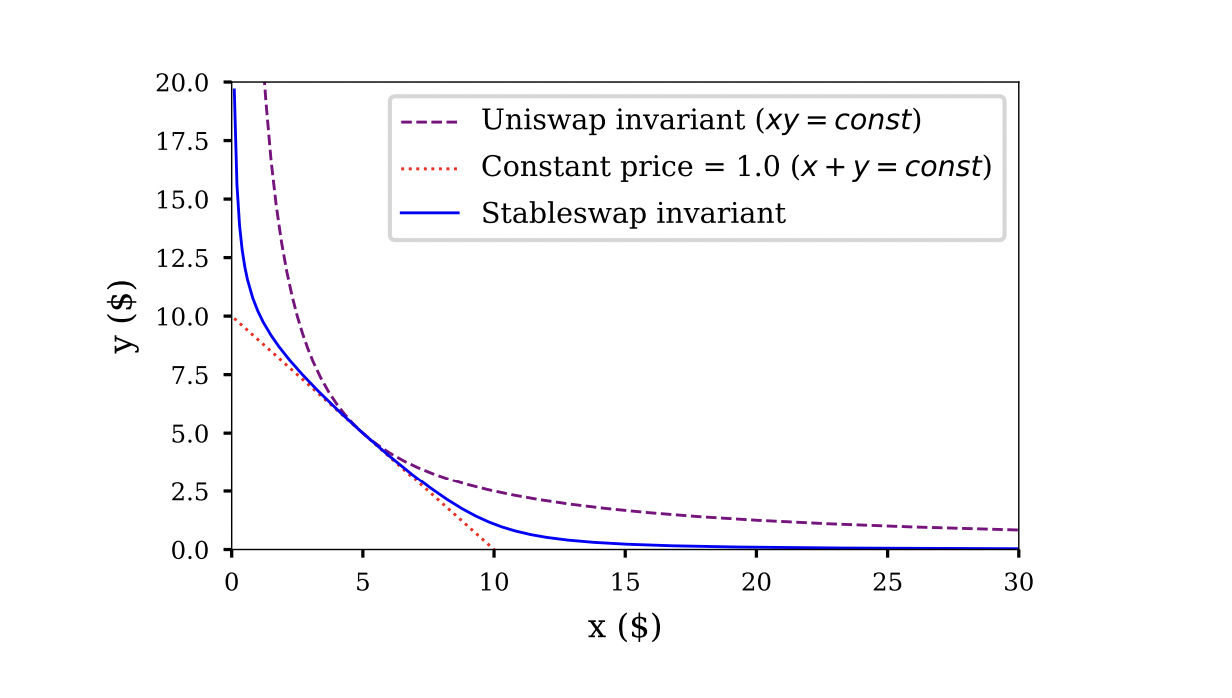}

\hypertarget{dynamic-chi}{%
\subsubsection{Dynamic Chi}\label{dynamic-chi}}

As described briefly in {[}Egorov19{]} and in more detail in {[}Feito{]}
the constant \(\chi\) (also called \(\chi\) in those papers, but our
\(k\) is their \(D\)) is not a constant but it is dynamic. The idea is
that the ideal state of the pool is to have the same number of both
tokens (and therefore the same value, as their natural price ratio is
unity). The curve as shown above has very little convexity in the
middle, and therefore very little slippage. There is very little
incentive for arbitrageurs to rebalance the pool, and it may remain off
kilter for a long time.

The stableswap mechanism therefore makes \(\chi\) dynamic: the further
the token ratio is away from unity, the smaller the \(\chi\), therefore
the closer the curve is to constant product, therefore the higher the
convexity, therefore the higher the slippage, and finally therefore the
higher the incentive for arbitrageurs to step in and balance the pool.

As \(\chi\) now depends on \(x\) in an implicit manner we can no longer
analytically calculate the price response function and the other
objects. However, intuitively we know how they look:

\begin{itemize}
\item
  the \textbf{price response function} \(\pi\) places most of the volume
  around unity price; however, when it gets close to the x axis it
  suddenly falls to zero, and when it gets to the right boundary it goes
  to infinity
\item
  the \textbf{portfolio value} function \(\nu\) is similar to that of
  the constant sum AMM, ie it ressembles an short put option profile
  that has been shifted upwards to go through the origin, with a bit of
  convexity added on either side
\item
  the \textbf{strikes} and the \textbf{Gamma} are placed away from the
  unity price point (not very far in absolute numbers, but very far away
  in terms of realistic movements)
\end{itemize}

\hypertarget{multi-asset}{%
\subsection{Multi asset}\label{multi-asset}}

\hypertarget{equal-weights}{%
\subsubsection{Equal weights}\label{equal-weights}}

As previously discussed, the characteristic function of a multi-asset
AMM is the product of the token amounts in native currencies. Therefore
the constant product AMM is a specific case of the multi-asset AMM,
albeit a rather special one as things get more complex in in higher
dimensions. The most commonly used function is the straight product

\[
\bar{k} = 
\bar f(x_0, x_1, \ldots, x_N) = 
x_0 \cdot x_1 \cdot \cdots \cdot x_N = 
\prod_{i=0}^N x_i
\]

As before it makes often sense to use a function that has scales
linearly, ie
\(f(\lambda x_0, \lambda x_1, \ldots) = \lambda f(x_0, x_1, \ldots)\)
because in this case the constant \(k\) is a measure of the pool size
that is not impacted by divergence loss. So instead of using the
straight product we are using the geometric average

\[
k = 
f(x_0, x_1, \ldots, x_N) = \sqrt[N+1]{x_0 \cdot x_1 \cdot \cdots \cdot x_N} =
\left(  \prod_{i=0}^N x_i\right  )^{\frac{1}{N+1}}
\]

The \textbf{indifference curve} is not an curve but a whole
\textbf{indifference surface}. We choose \(x_0\) as the numeraire that
we refer to as CSH -- a choice that is as arbitrary as choosing \(y\) in
the \(x*y\) case -- and the \(x_1\ldots x_N\) are the risk assets
RSK1\ldots RSKN. Isolating \(x_0\) we find

\[
x_{0;k}(x_1, x_2, \ldots, x_N) = 
\frac{k^{N+1}}{x_1\cdot x_2 \cdots x_N}
\]

In order to alleviate the notations we introduce the vector
\(x = (x_1, \ldots x_N)\), ie the vector of quantities of the risk
assets, but excluding the numeraire asset. Because we have multiple
assets we also now have multiple \textbf{price response functions}

\[
\pi_i(x) = 
-\frac{\partial x_0(x)}{\partial x_i} =
\frac{k^{N+1}}{x_1\cdots x_i^2\cdots x_N} =
\frac{x_0}{x_i}
\]

Note that we find the formula \(\pi=y/x\) from the two dimensional case,
and again the unit is \emph{CSH per RSKi}. In the above formula we find
partial derivatives, and they should not be looked at in isolation. They
should be considered a geometric object, notably the \emph{gradient
vector} corresponding to the indifference surface, ie the vector that is
orthogonal to its tangential plane. This plane has an important
financial interpretation: in the \(x*y\) case there is only one
direction in which one can move, so every trade of the risk asset RSK
forcibly involved the cash asset CSH. Now one can move inside this plane
without changing \emph{``height''}, ie without involving the cash asset.
This corresponds to direct trades between two risk assets RSKi and RSKj,
or more complex portfolios thereof when moving along a diagonal in the
tangent plane.

By definition, the AMM holdings of of RSKi are \(x_i\), and given the
price \(\pi_i\) above we find again that the AMM holds all assets in
equal value. In other words

\begin{quote}
In the unweighted multi-asset AMM in equilibrium with the market, the
monetary value of the CSH and all RSKi holdings is always equal
\end{quote}

We are now looking for the \textbf{normalized portfolio value} function
\(\nu(\xi)\) where \(\xi\) is the price ratio of asset RSKi compared to
time \(t=0\), and \(\nu\) is equally normalized to \(\nu(t=0)=1\). We
recall that for the constant product AMM we found that was
\(\nu(\xi)=\sqrt{\xi}\), which we proved by showing the hedging the
square root profile keeps half the value in the risk asset and half in
the numeraire asset. It is easy to verify that the function

\[
\nu(\xi_1 \ldots \xi_N) = \sqrt[N+1]{\xi_1 \cdots \xi_N} 
\]

satisfies this requirement. The calculation uses the fact that the cash
delta is calculated with the operator \(\xi_i \partial_i\) and that
\(\xi_i\partial_i(\xi_1 \cdots \xi_N)^\beta = \beta (\xi_1 \cdots \xi_N)^\beta\)
whatever \(\beta\), so if we choose \(\beta = \frac 1 {N+1}\) then each
of the \(N\) Cash Deltas is equal to \(\frac 1 {N+1}\) and together with
the value of the numeraire held the portfolio investment is distributed
evenly across all numeraire assets.

The HODL portfolio is initially equally invested in each of the N risk
assets as well as the numeraire, and at \(t=0\) we have \(\xi_i=1\), so
the value of the HODL portfolio is propotional to \(1 + \sum \xi_i\).
Propertly normalized the \textbf{divergence loss} is therefore

\[
\Lambda(\xi_1\ldots \xi_N) = \frac{1+\sum \xi_i}{N+1} - \sqrt[N+1]{\xi_1 \cdots \xi_N}
\]

\hypertarget{variable-weights}{%
\subsubsection{Variable weights}\label{variable-weights}}

We have previously looked at the case where all assets in the pool have
the same weight. Like in the two-dimensional case we can achieve
variable weights by introducing a coefficient vector
\(\alpha = \alpha_0, \ldots, \alpha_N\). We will assume that
\(\sum \alpha_i=1\) which ensures the homogenity of the function, ie we
get \(k\) instead of \(\bar k\). The \textbf{characteristic function} is
then

\[
k = f(x) = 
\prod_{i=0}^N x_i^{\alpha_i}
\]

Note that our \(\sqrt[N+1]{\ }\) term has been absorbed in the
\(\alpha\) which in the equally weighted case are equal to
\(\alpha_i = \frac 1 {N+1}\).

Similarly to the two-asset case we define

\[
\eta_i = \frac {\alpha_i} {\alpha_0} \Rightarrow 
\sum \eta_i = \frac{1-\alpha_0}{\alpha_0}, \ 
\frac 1 {\alpha_0} =  {1+\sum \eta_i}
\]

The \textbf{indifference surface} becomes

\[
x_{0;k}(x) = 
\left( \frac k {x_1^{\alpha_1} \cdots x_N^{\alpha_N}}  \right)^{\frac 1 {\alpha_0}} = 
\sqrt[1+\sum \eta_i] k \cdot \prod_{i=1}^N x_i^{-\eta_i}
\]

Note the minus sign in front of the exponent: the \(x\) are still in the
denominator but the formula becomes hard to read when writing it as a
fraction.

The \textbf{price response functions} becomes

\[
\pi_i(x) = 
-\frac{\partial x_0(x)}{\partial x_i} =
\eta_i \cdot \frac{x_0}{x_i} =
\frac{\alpha_i}{\alpha_0} \cdot \frac{x_0}{x_i} =
\frac{x_0/\alpha_0}{x_i/\alpha_i}
\]

ie it is adjusted with the relative weights factor \(\eta_i\). This also
implies that \(\eta_i\) is the relative weight of the assets in the
pool. In other words

\begin{quote}
In the variable weight multi-asset AMM in equilibrium with the market,
the monetary value of the RSKi holding is \(\eta_i\) times the CSH
holding where \(\eta_i = \alpha_i / \alpha_0\) is the ratio of the risk
asset and numeraire weight factors
\end{quote}

We should point out again that there is nothing special about the
numeraire asset, it was an arbitrary choice. So the above statement can
also be reformulated as

\begin{quote}
In the variable-weight multi-asset AMM in equilibrium with the market,
the relative monetary value of RSKi and RSKj holdings is
\(\alpha_i / \alpha_j\) (higher \(\alpha\) means bigger weight)
\end{quote}

The \textbf{normalized portfolio value} function in this case is

\[
\nu(\xi_1\ldots\xi_N) = 
\xi_1^{\alpha_1} \cdots \xi_N^{\alpha_N} 
= \prod_{i=1}^N \xi_i^{\alpha_i}
\]

It is easy to see that the Cash Delta is
\(\xi_i \partial_i \nu = \alpha_i \nu\) so indeed the portfolio
composition is in line with the coefficients \(\alpha_i\) and they sum
up to \(\nu\).

The HODL portfolio in this case is \(\alpha_0\) units for the numeraire
asset and \(\sum \alpha_i \xi_i\) for the risk assets, so the
\textbf{divergence loss} is

\[
\Lambda(\xi_1\ldots\xi_N) = 
\alpha_0 + \sum_{i=1}^{N} \alpha_i \xi_i 
-  \prod_{i=1}^N \xi_i^{\alpha_i}
\]

\hypertarget{conclusion}{%
\section{Conclusion}\label{conclusion}}

In this paper we have briefly reviewed the theory behind Automated
Market Makers, and we have provided specific formulas (also availble on
\href{https://theammbook.org/formulas}{theammbook.org/formulas} and
solutions for a number of the the important models in this space. Those
include the

\begin{itemize}
\item
  \textbf{\href{https://theammbook.org/formulas/constantproduct_cpmm/}{Constant
  Product}} designs like Bancor or Uniswap v2, and many other AMMs
  either on Ethereum or other chains who cover the full range of prices,
  which makes them versatile but capital inefficient even for regular
  tokens, and fully unsuitable for like-kind tokens (eg USDC vs DAI),
  the
\item
  \textbf{\href{https://theammbook.org/formulas/stableswap/}{Stableswap}}
  designs, created and popularized by Curve specifically for like-kind
  tokens for which the constant product design is highly inefficient,
  the
\item
  \textbf{\href{https://theammbook.org/formulas/varweight/}{Variable
  Weight}} designs that allow for a different portfolio composition than
  50:50 and that have advantages in some hub/spoke designs or for token
  distributors, the
\item
  \textbf{\href{https://theammbook.org/formulas/multiasset/}{Multi-Asset
  Pool}} designs like Balancer that allow for more capital efficient
  pools, variable weights, and one-hop trading across all tokens in the
  pool, and finally the
\item
  \textbf{\href{https://theammbook.org/formulas/concentrated/}{Concentrated
  Liquidity}} design of Uniswap v3 where liquidity provider are free to
  place their liquidity anywhere on the price curve, and where it is
  maximally levered, allowing to create arbitrary price response
  functions.
\end{itemize}

One class of AMMs we have deliberately excluded are any whose design
includes external data providers and oracles as those are in our view
very different designs that pose very different challenges. We plan to
cover those in a subsequent paper.

The world of AMMs is fast moving and we will keep this paper updated
with the important developments in this space. Please check on
\href{https://theammbook.org/paper}{theammbook.org/paper} for the most
recent version or - possibly a bit behind - on Arxiv once we have
finished the initial review cycle.

\hypertarget{references}{%
\section{References}\label{references}}

\begin{itemize}
\tightlist
\item
  \textbf{{[}BlackScholes73{]}} F Black, M Scholes: \emph{``The Pricing
  of Options and Corporate Liabilities''} (The Journal of Political
  Economy, 1973)
  \href{https://www.cs.princeton.edu/courses/archive/fall09/cos323/papers/black_scholes73.pdf}{url}
  \href{https://www.cs.princeton.edu/courses/archive/fall09/cos323/papers/black_scholes73.pdf}{pdf}
\item
  \textbf{{[}Merton73{]}} R Merton: \emph{``Theory of Rational Option
  Pricing''} (The Bell Journal of Economics and Management Science,
  1973) \href{https://www.jstor.org/stable/3003143}{url}
  \href{https://www.jstor.org/stable/3003143}{pdf}
\item
  \textbf{{[}Glosten94{]}} L Glosten: \emph{``Is the Electronic Open
  Limit Order Book Inevitable?''} (The Journal of Finance, 1994)
  \href{https://onlinelibrary.wiley.com/doi/abs/10.1111/j.1540-6261.1994.tb02450.x}{url}
  \href{https://onlinelibrary.wiley.com/doi/abs/10.1111/j.1540-6261.1994.tb02450.x}{pdf}
\item
  \textbf{{[}Derman99{]}} E Derman, K Demeterfi, M Kamal and J Zou:
  \emph{``A Guide to Volatility and Variance Swaps''} (The Journal of
  Derivatives, 1999)
  \href{https://jod.pm-research.com/content/6/4/9}{url}
  \href{https://jod.pm-research.com/content/6/4/9}{pdf}
\item
  \textbf{{[}Abernethy10{]}} J Abernethy, Y Chen, JW Vaughan: \emph{``An
  Optimization-Based Framework for Automated Market-Making''} (arXiv,
  2010) \href{https://arxiv.org/abs/1011.1941}{url}
  \href{https://arxiv.org/abs/1011.1941}{pdf}
\item
  \textbf{{[}Othman10{]}} A Othman, T Sandholm: \emph{``Automated
  market-making in the large: the gates hillman prediction market''}
  (Proceedings of the 11th ACM conference on Electronic commerce, 2010)
  \href{https://dl.acm.org/doi/10.1145/1807342.1807401}{url}
  \href{https://dl.acm.org/doi/10.1145/1807342.1807401}{pdf}
\item
  \textbf{{[}Othman12{]}} A Othman: \emph{``Automated Market Making:
  Theory and Practice (PhD Thesis)''} (Carnegy Mellon, 2012)
  \href{https://kilthub.cmu.edu/articles/thesis/Automated_Market_Making_Theory_and_Practice/6714920/1}{url}
  \href{https://kilthub.cmu.edu/ndownloader/files/12247904}{pdf}
\item
  \textbf{{[}Othman13{]}} A Othman, D Pennock, D Reeves, T Sandholm:
  \emph{``A Practical Liquidity-Sensitive Automated Market Maker''} (ACM
  Transactions on Economics and Computation, 2013)
  \href{https://www.cs.cmu.edu/~sandholm/liquidity-sensitive\%20automated\%20market\%20maker.teac.pdf}{url}
  \href{https://www.cs.cmu.edu/~sandholm/liquidity-sensitive\%20automated\%20market\%20maker.teac.pdf}{pdf}
\item
  \textbf{{[}Buterin17{]}} V Buterin: \emph{``On Path Independence''}
  (Private, 2017)
  \href{https://vitalik.ca/general/2017/06/22/marketmakers.html}{url}
  \href{https://vitalik.ca/general/2017/06/22/marketmakers.html}{pdf}
\item
  \textbf{{[}Buterin17b{]}} V Buterin: \emph{``Let's run on-chain
  decentralized exchanges the way we run prediction markets''} (Reddit,
  2017)
  \href{https://www.reddit.com/r/ethereum/comments/55m04x/lets_run_onchain_decentralized_exchanges_the_way/}{url}
  \href{https://www.reddit.com/r/ethereum/comments/55m04x/lets_run_onchain_decentralized_exchanges_the_way/}{pdf}
\item
  \textbf{{[}Hertzog17{]}} E Hertzog, G Benartzi, G Benartzi:
  \emph{``Continuous Liquidity and Asynchronous Price Discovery for
  Tokens through their Smart Contracts; aka''Smart Tokens''"} (Bancor,
  2017) \href{https://whitepaper.io/document/52/bancor-whitepaper}{url}
  \href{https://whitepaper.io/document/52/bancor-whitepaper}{pdf}
\item
  \textbf{{[}Lu17{]}} A Lu: \emph{``Building a Decentralized Exchange in
  Ethereum''} (Medium, 2017)
  \href{https://blog.gnosis.pm/building-a-decentralized-exchange-in-ethereum-eea4e7452d6e}{url}
  \href{https://blog.gnosis.pm/building-a-decentralized-exchange-in-ethereum-eea4e7452d6e}{pdf}
\item
  \textbf{{[}Buterin18{]}} V Buterin: \_"Improving front running
  resistance of x*y=k market makers"\_ (Private, 2018)
  \href{https://ethresear.ch/t/improving-front-running-resistance-of-x-y-k-market-makers/1281}{url}
  \href{https://ethresear.ch/t/improving-front-running-resistance-of-x-y-k-market-makers/1281}{pdf}
\item
  \textbf{{[}Hertzog18{]}} E Hertzog, G Benartzi, G Benartzi, O Ross:
  \emph{``Bancor Protocol - Continuous Liquidity for Cryptographic
  Tokens through their Smart Contracts''} (Bancor, 2018)
  \href{https://storage.googleapis.com/website-bancor/2018/04/01ba8253-bancor_protocol_whitepaper_en.pdf}{url}
  \href{https://storage.googleapis.com/website-bancor/2018/04/01ba8253-bancor_protocol_whitepaper_en.pdf}{pdf}
\item
  \textbf{{[}Zargham18{]}} M Zargham, Z Zhang, V Preciado: \emph{``A
  State-Space Modeling Framework for Engineering Blockchain-Enabled
  Economic Systems''} (Arxiv, 2018)
  \href{https://arxiv.org/abs/1807.00955}{url}
  \href{https://arxiv.org/abs/1807.00955}{pdf}
\item
  \textbf{{[}Adams19{]}} H Adams: \emph{``Uniswap Birthday Blog ---
  V0''} (Uniswap, 2019)
  \href{https://medium.com/uniswap/uniswap-birthday-blog-v0-7a91f3f6a1ba}{url}
  \href{https://medium.com/uniswap/uniswap-birthday-blog-v0-7a91f3f6a1ba}{pdf}
\item
  \textbf{{[}Daian19{]}} P Daian, S Goldfeder, T Kell, Y Li, X Zhao, I
  Bentov, L Breidenbach, A Juels: \emph{``Flash Boys 2.0: Frontrunning,
  Transaction Reordering, and Consensus Instability in Decentralized
  Exchanges''} (Arxiv, 2019)
  \href{https://arxiv.org/abs/1904.05234}{url}
  \href{https://arxiv.org/abs/1904.05234}{pdf}
\item
  \textbf{{[}DiMaggio19{]}} M Di Maggio: \emph{``Survey of Automated
  Market Making Algorithms''} (Medium, 2019)
  \href{https://medium.com/terra-money/survey-of-automated-market-making-algorithms-951f91ce727a}{url}
  \href{https://medium.com/terra-money/survey-of-automated-market-making-algorithms-951f91ce727a}{pdf}
\item
  \textbf{{[}Egorov19{]}} M Egorov: \emph{``StableSwap - a efficient
  mechanism for Stablecoin liquidity''} (Curve, 2019)
  \href{https://curve.fi/files/stableswap-paper.pdf}{url}
  \href{https://curve.fi/files/stableswap-paper.pdf}{pdf}
\item
  \textbf{{[}Kereiakes19{]}} E Kereiakes, D Kwon, M Di Maggio, N
  Platias: \emph{``Terra Money: Stability and Adoption''} (Terra, 2019)
  \href{https://assets.website-files.com/611153e7af981472d8da199c/618b02d13e938ae1f8ad1e45_Terra_White_paper.pdf}{url}
  \href{https://assets.website-files.com/611153e7af981472d8da199c/618b02d13e938ae1f8ad1e45_Terra_White_paper.pdf}{pdf}
\item
  \textbf{{[}Martinelli19{]}} F Martinelli, N Mushegian:
  \emph{``Balancer Whitepaper''} (Balancer, 2019)
  \href{https://balancer.fi/whitepaper.pdf}{url}
  \href{https://balancer.fi/whitepaper.pdf}{pdf}
\item
  \textbf{{[}Pintail19{]}} Pintail: \emph{``Uniswap: A Good Deal for
  Liquidity Providers?''} (Medium, 2019)
  \href{https://pintail.medium.com/uniswap-a-good-deal-for-liquidity-providers-104c0b6816f2}{url}
  \href{https://pintail.medium.com/uniswap-a-good-deal-for-liquidity-providers-104c0b6816f2}{pdf}
\item
  \textbf{{[}Angeris20{]}} G Angeris, A Evans, T Chitra: \emph{``When
  does the tail wag the dog? Curvature and market making''} (Arxiv,
  2020) \href{https://arxiv.org/pdf/2012.08040}{url}
  \href{https://arxiv.org/pdf/2012.08040.pdf}{pdf}
\item
  \textbf{{[}Angeris20a{]}} G Angeris, T Chitra: \emph{``Improved Price
  Oracles''} (Proceedings of the 2nd ACM Conference on Advances in
  Financial Technologies, 2020)
  \href{http://dx.doi.org/10.1145/3419614.3423251}{url}
  \href{https://arxiv.org/pdf/2003.10001.pdf}{pdf}
\item
  \textbf{{[}Bukov20{]}} A Bukov, M Melnik: \emph{``Mooniswap by 1inch
  exchange)''} (1inch, 2020)
  \href{https://mooniswap.exchange/docs/MooniswapWhitePaper-v1.0.pdf}{url}
  \href{https://mooniswap.exchange/docs/MooniswapWhitePaper-v1.0.pdf}{pdf}
\item
  \textbf{{[}Clark20{]}} J Clark: \emph{``The Replicating Portfolio of a
  Constant Product Market''} (SSRN, 2020)
  \href{https://papers.ssrn.com/sol3/papers.cfm?abstract_id=3550601}{url}
  \href{https://papers.ssrn.com/sol3/Delivery.cfm/SSRN_ID3550601_code2295282.pdf?abstractid=3550601\&mirid=1}{pdf}
\item
  \textbf{{[}Dodo20{]}} DODO Team: \emph{``Why There is No Impermanent
  Loss on DODO''} (DODOex, 2020)
  \href{https://blog.dodoex.io/why-there-is-no-impermanent-loss-on-dodo-4d82b2b532bc}{url}
  \href{https://blog.dodoex.io/why-there-is-no-impermanent-loss-on-dodo-4d82b2b532bc}{pdf}
\item
  \textbf{{[}Dodo20a{]}} DODO Team: \emph{``Introducing DODO: 10x Better
  Liquidity than Uniswap''} (DODOex, 2020)
  \href{https://dodo-in-the-zoo.medium.com/introducing-dodo-10x-better-liquidity-than-uniswap-852ce2137c57}{url}
  \href{https://dodo-in-the-zoo.medium.com/introducing-dodo-10x-better-liquidity-than-uniswap-852ce2137c57}{pdf}
\item
  \textbf{{[}Feito20{]}} A Feito: \emph{``A stablecoin exchange''}
  (Curve, 2020) \href{https://alvarofeito.com/articles/curve/}{url}
  \href{https://alvarofeito.com/articles/curve/}{pdf}
\item
  \textbf{{[}KPMG20{]}} KPMG: \emph{``Automated Market Makers -
  Innovations, Challenges, Prospects''} (KPMG, 2020)
  \href{https://assets.kpmg/content/dam/kpmg/cn/pdf/en/2021/10/crypto-insights-part-2-decentralised-exchanges-and-automated-market-makers.pdf}{url}
  \href{https://assets.kpmg/content/dam/kpmg/cn/pdf/en/2021/10/crypto-insights-part-2-decentralised-exchanges-and-automated-market-makers.pdf}{pdf}
\item
  \textbf{{[}Loesch20{]}} S Loesch, N Hindman: \emph{``Bancor Protocol
  v2.1 - Economic and Quantitative Finance Analysis''} (Private, 2020)
  \href{https://drive.google.com/file/d/1en044m2wchn85aQBcoVx2elmxEYd5kEA/view}{url}
  \href{https://drive.google.com/file/d/1en044m2wchn85aQBcoVx2elmxEYd5kEA/view}{pdf}
\item
  \textbf{{[}Lyons20{]}} R Lyons, G Viswanath-Natraj: \emph{``What keeps
  stable coins stable?''} (NBER, 2020)
  \href{https://www.nber.org/system/files/working_papers/w27136/w27136.pdf}{url}
  \href{https://www.nber.org/system/files/working_papers/w27136/w27136.pdf}{pdf}
\item
  \textbf{{[}Michael20{]}} Michael: \emph{``Amplified Liquidity:
  Designing Capital Efficient Automated Market Makers in Bancor V2''}
  (Bancor, 2020)
  \href{https://blog.bancor.network/amplified-liquidity-designing-capital-efficient-automated-market-makers-in-bancor-v2-3cec8891c3a1}{url}
  \href{https://blog.bancor.network/amplified-liquidity-designing-capital-efficient-automated-market-makers-in-bancor-v2-3cec8891c3a1}{pdf}
\item
  \textbf{{[}Niemerg20{]}} A Niemerg, D Robinson, L Livnev:
  \emph{``YieldSpace: An Automated Liquidity Provider for Fixed Yield
  Tokens''} (Yield, 2020) \href{https://yield.is/YieldSpace.pdf}{url}
  \href{https://yield.is/YieldSpace.pdf}{pdf}
\item
  \textbf{{[}Noyes20{]}} C Noyes: \emph{``Liquidity Provider Wealth''}
  (Paradigm, 2020)
  \href{https://research.paradigm.xyz/LP_Wealth.pdf}{url}
  \href{https://research.paradigm.xyz/LP_Wealth.pdf}{pdf}
\item
  \textbf{{[}THORChain20{]}} THORChain Team: \emph{``THORChain - A
  decentralized liquidity network''} (THOrChain, 2020)
  \href{https://github.com/thorchain/Resources/blob/master/Whitepapers/THORChain-Whitepaper-May2020.pdf}{url}
  \href{https://github.com/thorchain/Resources/blob/master/Whitepapers/THORChain-Whitepaper-May2020.pdf}{pdf}
\item
  \textbf{{[}Tassy20{]}} M Tassy, D White: \emph{``Growth rate of a
  liquidity provider's wealth in xy=c automated market makers''}
  (Private, 2020)
  \href{https://math.dartmouth.edu/~mtassy/articles/AMM_returns.pdf}{url}
  \href{https://math.dartmouth.edu/~mtassy/articles/AMM_returns.pdf}{pdf}
\item
  \textbf{{[}Zhang20{]}} D Zhang: \emph{``Curve.fi 101 --- How it works
  and its meteoric rise''} (Private, 2020)
  \href{https://medium.com/stably-blog/curve-fi-101-how-it-works-and-its-meteoric-rise-2d6dffd30d51}{url}
  \href{https://medium.com/stably-blog/curve-fi-101-how-it-works-and-its-meteoric-rise-2d6dffd30d51}{pdf}
\item
  \textbf{{[}Adams21{]}} Hayden Adams and Noah Zinsmeister and Moody
  Salem and River Keefer and Dan Robinson: \emph{``Uniswap v3 Core''}
  (Uniswap, 2021) \href{https://uniswap.org/whitepaper-v3.pdf}{url}
  \href{https://uniswap.org/whitepaper-v3.pdf}{pdf}
\item
  \textbf{{[}Angeris21{]}} G Angeris, A Agrawal, A Evans, T Chitra, S
  Boyd: \emph{``Constant Function Market Makers: Multi-Asset Trades via
  Convex Optimization''} (Arxiv, 2021)
  \href{https://arxiv.org/pdf/2107.12484}{url}
  \href{https://arxiv.org/pdf/2107.12484.pdf}{pdf}
\item
  \textbf{{[}Angeris21a{]}} G Angeris, A Evans, T Chitra:
  \emph{``Replicating Market Makers''} (Arxiv, 2021)
  \href{https://arxiv.org/pdf/2103.14769}{url}
  \href{https://arxiv.org/pdf/2103.14769.pdf}{pdf}
\item
  \textbf{{[}Angeris21b{]}} G Angeris, A Evans, T Chitra: \emph{``A Note
  on Privacy in Constant Function Market Makers''} (Arxiv, 2021)
  \href{https://arxiv.org/pdf/2103.01193}{url}
  \href{https://arxiv.org/pdf/2103.01193.pdf}{pdf}
\item
  \textbf{{[}Angeris21c{]}} G Angeris, H Kao, R Chiang, C Noyes, T
  Chitra: \emph{``An analysis of Uniswap markets''} (Arxiv, 2021)
  \href{https://arxiv.org/pdf/1911.03380}{url}
  \href{https://arxiv.org/pdf/1911.03380.pdf}{pdf}
\item
  \textbf{{[}Angeris21d{]}} G Angeris, A Evans, T Chitra:
  \emph{``Replicating Monotonic Payoffs Without Oracles''} (Arxiv, 2021)
  \href{https://arxiv.org/pdf/2111.13740}{url}
  \href{https://arxiv.org/pdf/2111.13740.pdf}{pdf}
\item
  \textbf{{[}Clark21{]}} J Clark: \emph{``A deconstructed constant
  product market''} (SSRN, 2021)
  \href{https://papers.ssrn.com/sol3/papers.cfm?abstract_id=3856446}{url}
  \href{https://papers.ssrn.com/sol3/papers.cfm?abstract_id=3856446}{pdf}
\item
  \textbf{{[}Dodo21{]}} DODO Team: \emph{``DODO - A Next-Generation
  On-Chain Liquidity Provider Powered by Pro-active Market Maker
  Algorithm''} (DODOex, 2021)
  \href{https://dodoex.github.io/docs/docs/whitepaper/}{url}
  \href{https://dodoex.github.io/docs/docs/whitepaper/}{pdf}
\item
  \textbf{{[}Egorov21{]}} M Egorov: \emph{``Automatic market-making with
  dynamic peg''} (Curve, 2021)
  \href{https://curve.fi/files/crypto-pools-paper.pdf}{url}
  \href{https://curve.fi/files/crypto-pools-paper.pdf}{pdf}
\item
  \textbf{{[}Evans21{]}} A Evans, G Angeris, T Chitra: \emph{``Optimal
  Fees for Geometric Mean Market Makers''} (Arxiv, 2021)
  \href{https://arxiv.org/pdf/2104.00446}{url}
  \href{https://arxiv.org/pdf/2104.00446.pdf}{pdf}
\item
  \textbf{{[}Falakshahi21{]}} H Falakshahi, M Mariapragassam, R Ajaja:
  \emph{``Automated Market Making with Synchronized Liquidity Pools''}
  (SSRN, 2021)
  \href{https://papers.ssrn.com/sol3/papers.cfm?abstract_id=3963811}{url}
  \href{https://papers.ssrn.com/sol3/papers.cfm?abstract_id=3963811}{pdf}
\item
  \textbf{{[}Grove21{]}} C Grove: \emph{``Price Formulation of CFMMs''}
  (Private, 2021)
  \href{https://github.com/galacticcouncil/Research/blob/main/PriceFormulationOfAMMs/PriceFormulationOfAMMs.pdf}{url}
  \href{https://github.com/galacticcouncil/Research/blob/main/PriceFormulationOfAMMs/PriceFormulationOfAMMs.pdf}{pdf}
\item
  \textbf{{[}Lambert21{]}} G Lambert: \emph{``Understanding the Value of
  Uniswap v3 Liquidity Positions''} (Medium, 2021)
  \href{https://lambert-guillaume.medium.com/understanding-the-value-of-uniswap-v3-liquidity-positions-cdaaee127fe7}{url}
  \href{https://lambert-guillaume.medium.com/understanding-the-value-of-uniswap-v3-liquidity-positions-cdaaee127fe7}{pdf}
\item
  \textbf{{[}Loesch21{]}} S Loesch, N Hindman, MB Richardson, N Welch:
  \emph{``Impermanent Loss in Uniswap v3''} (Arxiv, 2021)
  \href{https://arxiv.org/pdf/2111.09192.pdf}{url}
  \href{https://arxiv.org/pdf/2111.09192.pdf}{pdf}
\item
  \textbf{{[}Mota21{]}} M Mota: \emph{``Understanding StableSwap''}
  (Private, 2021)
  \href{https://miguelmota.com/blog/understanding-stableswap-curve/}{url}
  \href{https://miguelmota.com/blog/understanding-stableswap-curve/}{pdf}
\item
  \textbf{{[}Nguyen21{]}} A Nguyen, L Luu, M Ng: \emph{``Dynamic
  Automated Market Making''} (Kyber, 2021)
  \href{https://files.kyber.network/DMM-Feb21.pdf}{url}
  \href{https://files.kyber.network/DMM-Feb21.pdf}{pdf}
\item
  \textbf{{[}Onomy21{]}} Onomy Protocol: \emph{``An Analysis of
  Automated Market Makers vs.~Order Books''} (Onomy, 2021)
  \href{https://medium.com/onomy-protocol/an-analysis-of-automated-market-makers-vs-order-books-f1af65200ca6}{url}
  \href{https://medium.com/onomy-protocol/an-analysis-of-automated-market-makers-vs-order-books-f1af65200ca6}{pdf}
\item
  \textbf{{[}Platypus21{]}} Platypus Team: \emph{``Platypus AMM
  Technical Specification''} (Platypus, 2021)
  \href{https://drive.google.com/file/d/11qScw60LAb83_zq-E5ttolaIbeg5jQ0z/view}{url}
  \href{https://drive.google.com/file/d/11qScw60LAb83_zq-E5ttolaIbeg5jQ0z/view}{pdf}
\item
  \textbf{{[}Tassy21{]}} M Tassy: \emph{``A different view at AMMs''}
  (Private, 2021)
  \href{https://martintassy.wordpress.com/2022/01/03/a-different-view-at-amms}{url}
  \href{https://martintassy.wordpress.com/2022/01/03/a-different-view-at-amms}{pdf}
\item
  \textbf{{[}Tassy21a{]}} M Tassy: \emph{``From Uniswap to option AMMs
  (Part I): A different view at AMMs''} (Private, 2021)
  \href{https://martintassy.wordpress.com/2021/08/01/from-uniswap-to-option-amms-part-i-a-different-view-at-amms/}{url}
  \href{https://martintassy.wordpress.com/2021/08/01/from-uniswap-to-option-amms-part-i-a-different-view-at-amms/}{pdf}
\item
  \textbf{{[}Lambert22{]}} G Lambert: \emph{``How to deploy
  delta-neutral liquidity in Uniswap''} (Medium, 2022)
  \href{https://lambert-guillaume.medium.com/how-to-deploy-delta-neutral-liquidity-in-uniswap-or-why-euler-finance-is-a-game-changer-for-lps-1d91efe1e8ac}{url}
  \href{https://lambert-guillaume.medium.com/how-to-deploy-delta-neutral-liquidity-in-uniswap-or-why-euler-finance-is-a-game-changer-for-lps-1d91efe1e8ac}{pdf}
\item
  \textbf{{[}Element22{]}} Element: \emph{``The Element Protocol
  Construction Paper''} (Element, Undated)
  \href{https://paper.element.fi/}{url}
  \href{https://paper.element.fi/}{pdf}
\item
  \textbf{{[}Hull{]}} J Hull: \emph{``Options, Futures, and Other
  Derivatives''} (Pearson, 2018)
  \href{https://www.pearson.com/us/higher-education/program/Hull-Options-Futures-and-Other-Derivatives-10th-Edition/PGM333301.html}{url}
  \href{https://www.pearson.com/us/higher-education/program/Hull-Options-Futures-and-Other-Derivatives-10th-Edition/PGM333301.html}{pdf}
\item
  \textbf{{[}Loesch{]}} S Loesch: \emph{``A Guide to Financial
  Regulation for Fintech Entrepreneurs''} (Wiley, 2018)
  \href{https://onlinelibrary.wiley.com/doi/book/10.1002/9781119436775}{url}
  \href{https://onlinelibrary.wiley.com/doi/book/10.1002/9781119436775}{pdf}
\item
  \textbf{{[}Pyndick-Rubinfeld{]}} R Pyndick, D Rubinfeld:
  \emph{``Microeconomics''} (Pearson, 2018)
  \href{https://www.pearson.com/uk/educators/higher-education-educators/program/Pindyck-Microeconomics-Global-Edition-9th-Edition/PGM1825855.html}{url}
  \href{https://www.pearson.com/uk/educators/higher-education-educators/program/Pindyck-Microeconomics-Global-Edition-9th-Edition/PGM1825855.html}{pdf}
\end{itemize}

\hypertarget{appendix}{%
\section{Appendix}\label{appendix}}

\hypertarget{website}{%
\subsection{Website}\label{website}}

The website of the book is at
\href{https://theammbook.org}{theammbook.org}. The paper itself can be
found on the website at
\href{https://theammbook.org/paper}{theammbook.org/paper}

\hypertarget{glossary-and-technical-glossary}{%
\subsection{Glossary and technical
glossary}\label{glossary-and-technical-glossary}}

A glossary of general terms is at
\href{https://theammbook.org/glossary}{theammbook.org/glossary}. A
technical (formula) glossary is at
\href{https://theammbook.org/techglossary}{theammbook.org/techglossary}

\hypertarget{formulas}{%
\subsection{Formulas}\label{formulas}}

Key AMM-related formulas are are
\href{https://theammbook.org/formulas}{theammbook.org/formulas}

\hypertarget{projects}{%
\subsection{Projects}\label{projects}}

A list of AMM and AMM-related projects is at
\href{https://theammbook.org/projects}{theammbook.org/projects}

\hypertarget{references-1}{%
\subsection{References}\label{references-1}}

A list of AMM-related references (academic papers, blogs, books) is at
\href{https://theammbook.org/references}{theammbook.org/references}

\hypertarget{license}{%
\section{License}\label{license}}

This work is licensed under a
\href{https://creativecommons.org/licenses/by/4.0/}{Creative Commons
Attribution 4.0 International License}.

\includegraphics[width=2cm,keepaspectratio]{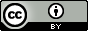}

\hypertarget{attribution-4.0-international-cc-by-4.0}{%
\subsection{Attribution 4.0 International (CC BY
4.0)}\label{attribution-4.0-international-cc-by-4.0}}

This is a human-readable summary of (and not a substitute for) the
\href{https://creativecommons.org/licenses/by/4.0/legalcode}{license}.

\begin{quote}
This deed highlights only some of the key features and terms of the
actual license. It is not a license and has no legal value. You should
carefully review all of the terms and conditions of the
\href{https://creativecommons.org/licenses/by/4.0/legalcode}{actual
license} before using the licensed material.
\end{quote}

\textbf{You are free to:}

\begin{itemize}
\item
  \textbf{Share} --- copy and redistribute the material in any medium or
  format
\item
  \textbf{Adapt} --- remix, transform, and build upon the material
\end{itemize}

for any purpose, even commercially. The licensor cannot revoke these
freedoms as long as you follow the license terms.

\textbf{Under the following terms:}

\begin{itemize}
\item
  \textbf{Attribution} --- You must give appropriate credit, provide a
  link to the license, and indicate if changes were made. You may do so
  in any reasonable manner, but not in any way that suggests the
  licensor endorses you or your use.
\item
  \textbf{No additional restrictions} --- You may not apply legal terms
  or technological measures that legally restrict others from doing
  anything the license permits.
\end{itemize}

\textbf{Notices:}

You do not have to comply with the license for elements of the material
in the public domain or where your use is permitted by an applicable
exception or limitation. No warranties are given. The license may not
give you all of the permissions necessary for your intended use. For
example, other rights such as publicity, privacy, or moral rights may
limit how you use the material.

\end{document}